\documentclass[reprint, amsmath, amssymb, aps, prd, nofootinbib]{revtex4-1}

\usepackage{graphicx}

\usepackage{bm}

\usepackage{hyperref}

\usepackage{physics}

\usepackage[caption=false]{subfig}

\usepackage{orcidlink}

\def\apj{\rm{ApJ}}
\def\apjl{\rm{ApJ}}
\def\apjs{\rm{ApJS}}
\def\aap{\rm{A\&A}}

\def\mnras{\rm{MNRAS}}
\def\prd{\rm{PRD}}
\def\prl{\rm{PRL}}

\newcommand{\bham}{Institute for Gravitational Wave Astronomy \& School of Physics and Astronomy, University of Birmingham, Edgbaston, Birmingham B15 2TT, UK}

\begin{document}

\title{
LISA stellar-mass black hole searches with semicoherent and particle-swarm methods
}

\author{Diganta Bandopadhyay
\orcidlink{0000-0003-0975-5613}}
\email{diganta@star.sr.bham.ac.uk}
\affiliation{\bham}

\author{Christopher J. Moore
\orcidlink{0000-0002-2527-0213}}
\email{cmoore@star.sr.bham.ac.uk}
\affiliation{\bham}

\date{\today}

\begin{abstract}
    This paper considers the problem of searching for quiet, long-duration and broadband gravitational wave signals, such as stellar-mass binary black hole binaries, in mock LISA data.
    We propose a method that combines a semi-coherent likelihood with the use of a particle swarm optimizer capable of efficiently exploring a large parameter space. 
    The semi-coherent analysis is used to widen the peak of the likelihood distribution over parameter space, congealing secondary peaks and thereby assisting in localizing the posterior bulk. 
    An iterative strategy is proposed, using particle swarm methods to initially explore a wide, loosely-coherent likelihood and then progressively constraining the signal to smaller regions in parameter space by increasing the level of coherence. 
    The properties of the semi-coherent likelihood are first demonstrated using the well-studied binary neutron star signal GW170817.
    As a proof of concept, the method is then successfully applied to a simplified search for a stellar-mass binary black hole in zero-noise LISA data. 
    Finally, we conclude by discussing what remains to be done to develop this into a fully-capable search and how the method might also be adapted to tackle the EMRI search problem in LISA. 
\end{abstract}

\maketitle

\section{Introduction}\label{sec:introduction}

The Laser Interferometer Space Antenna (LISA) \cite{Seoane_2017_LISA} will detect low-frequency ($\sim 0.1\textrm{\textendash}100\,\mathrm{mHz}$) gravitational waves (GWs) from a wide range of astrophysical sources.
Among these, stellar-mass binary black hole (SmBBH) \citep{2010_Amaro_Seoana_SmBBH,2016_Sesana_multimessenger} and extreme-mass-ratio inspiral (EMRI) \cite{Babak_EMRI_2017, 2007CQGra..24R.113A, Gair_2017_EMRI_prospects} sources are of particular interest here. 
SmBBHs consist of a pair of approximately equal-mass black holes (BHs) in the mass range $\sim 10\textrm{\textendash}100\,M_{\odot}$, and LISA will observe many of these $\sim 1\textrm{\textendash}10$ years before merger.
In contrast, EMRIs consist of a supermassive BH, with mass in the range $\sim 10^4 \textrm{\textendash} 10^7 M_{\odot}$ as found in the centers of most galaxies, orbited by a stellar-mass compact object with a mass in the range $\sim 1\textrm{\textendash}100\,M_{\odot}$. 
SmBBHs will eventually merge in the LIGO/Virgo \cite{LIGO_ref,LIGO_ref_2} frequency band; events similar to GW150914 \cite{Abbott_2016_150914} and GW190521 \cite{Abbott_2020_190521} would have appeared as quiet, long-lived LISA sources, had the instrument been operating several years previously.

SmBBH systems are observed by LISA relatively early in their inspiral, when the orbital separation is much larger than the Schwarzschild radius of either BH.
At this stage in the inspiral the orbital velocity is small, $v\ll c$, and these systems are weak sources of GWs with the GW frequency evolving slowly (i.e.\ approximately adiabatically) and the source completing many orbits in the LISA frequency band.
EMRI signals meanwhile are observed late in their inspiral, when the orbital separation is comparable to the radius of the larger BH.
These are highly relativistic sources with $v\lesssim c$.
However, the extreme mass-ratio of these systems means that they are also weak sources of GWs with slowly evolving frequencies.
Again, this leads to a large number of orbits being completed in the LISA band. 
Although SmBBHs and EMRIs are physically very different, both will appear as long-lived, broadband signals in LISA with $\gtrsim 10^{5}$ observable GW cycles.
From a data analysis perspective, the main difference between the two source types is that SmBBH signals are dominated by a single-frequency harmonic, whereas EMRI signals may have significant contributions from many harmonics. SmBBHs and EMRIs promise exciting new possibilities for multimessenger \cite{2022_Klein_Multiband,2016_Sesana_multimessenger} and fundamental physics \cite{2019_Berry_EMRI_motivation}. 

The problem of detecting and characterizing a GW source involves finding the waveform models and parameters that best fit the observed data. 
This process is conventionally split into two phases: \textit{search} and \textit{parameter estimation}.
The search phase aims to identify if the data contains a source (or sources) and its approximate parameters. This will be extremely challenging for SmBBH and EMRI signals due to the size of the parameter space that must be explored. Ideas for EMRI search strategies have been investigated in Refs.~\citep{2009_Babak_EMRI_search,2011_Cornish_EMRI_search,2005_Gair_EMRI_search,2005_Wen_EMRI_search,2007_Gair_EMRI_search}.
Once a search identifies a candidate detection, the parameter estimation phase is tackled using a well-established Bayesian framework that maps out the posterior distribution on the waveform parameters.
Parameter estimation for both EMRI \cite{2021_Katz_FastEMRIwaveforms,2023_Speri_FD_EMRI} and SmBBH \citep{2022_Klein_Multiband,Toubiana_2020_SmBBH,Digman_2022_SmBBH,Buscicchio_2021,Sberna_2022_190521} signals has been previously demonstrated. 
The holy grail of source characterization for LISA is the global fit which aims to simultaneously estimate the source parameters of all the signals observed by LISA \cite{2005_Cornish_Global_fit}, a recent prototype implementation is shown in Ref.~\cite{Tyson_2024_Global_Fit}. 
Sources will be chirping and overlapping in both time and frequency; disentangling each source from the combined data-stream will be an extremely challenging problem. 
The global fit is made tractable via the prior identification of regions in parameter space where signals might exist; this prior identification is the role of the search phase.  
This primary search phase is an open problem for SmBBH and EMRI signals \citep{Moore_2019_SmBBH_template_bank,Gair_2004_EMRI_template_bank} and is the subject of this paper.

Long-lived signals, such as SmBBH and EMRIs, undergo a large number of orbits. This allows certain parameters that control the GW frequency (notably the binary chirp mass, $\mathcal{M}_c$) to be measured with exquisite precision.
Among the current GW detections, the closest analog to an SmBBH or EMRI signal is the binary neutron star (BNS) GW170817 \cite{Abbot_2017_BNS}. 
This low-mass ($\mathcal{M}_c\sim 1\,M_\odot$) source completed $\sim 3000$ cycles in the detector frequency band, compared to just $\sim10$ for the high-mass ($\mathcal{M}_c\sim 30\,M_\odot$) GW150914 binary BH \cite{Abbott_2016_150914}.
The longer signal translates to a more precise measurement of the chirp mass; for GW170817 the fractional error was $\delta \mathcal{M}_c/\mathcal{M}_c\sim 10^{-3}$ \cite{Abbot_2017_BNS} whereas for the much shorter GW150914 it was $\sim 10^{-1}$.
In contrast, for the extremely long SmBBH and EMRI systems observed in LISA, the fractional error on the chirp mass is expected to be several orders of magnitude smaller, depending on the source parameters; for example, in the case of a GW190521-like source observable by LISA, the fractional uncertainty is predicted to be $10^{-5}$ \citep{Sberna_2022_190521,2022_Klein_Multiband}. The precision of these measurements drives the requirements for the search; for systems where we can measure the source parameters with greater precision, the search must cover the parameter space with a correspondingly finer resolution. SmBBH and EMRI signals in LISA represent a completely new challenge, orders of magnitude more difficult than those encountered to date in GW astronomy. This calls for completely new analysis tools and methods.

Searches for compact binary coalescences in LIGO-Virgo data have been successfully conducted using template banks since the very first detection \cite{Abbott_2016_template_banks}. 
A template bank comprises a set of model waveforms, known as templates, evaluated at a predetermined set of locations in parameter space. 
A search matches the data against each template in the bank; if the template with the highest match passes some threshold, a detection is claimed and the parameters of this template are then used to inform subsequent parameter estimation. 
The template bank used for the detection of GW170817 was much denser (in the sense that the spacing of templates in, say, chirp mass was smaller) than that for the shorter GW150914 signal \cite{Roulet_2019_170817_template_bank}.  
Estimates suggest that a template-based search for EMRI signals, with orders of magnitude more cycles in band, would require $\sim 10^{40}$ templates to cover the parameter space \cite{Gair_2004_EMRI_template_bank}, rendering the approach unfeasible. 
Template bank searches for SmBBHs suffer a similar problem, albeit to a somewhat lesser degree \cite{Moore_2019_SmBBH_template_bank}.

It is interesting to note that, if one was prepared to wait several years, one could rely on some future-generation ground-based detector to observe the final merger of the SmBBH systems. 
This could be used as a trigger to go back and perform parameter estimation on the archival LISA data without the need for a full search.
Such archival searches have been demonstrated for quiet SmBBH mock LISA signals \cite{Ewing_2021_archival}. It is worth noting that even in the archival targeted search scenario, localizing a signal in the constrained parameter space can still be challenging \cite{2023_Wang_SmBBH_template_bank}. We do not want to rely solely on archival searches for several reasons: ground-based detectors will not operate with a 100\% duty cycle and will therefore miss a fraction of events; the prospect of an advanced warning of a GW event alongside a sky localization can be invaluable for multimessenger astronomy \citep{2022_Klein_Multiband,2016_Sesana_multimessenger,2017_Sesana_multimessenger}; and archival searches will not be possible at all for EMRIs.

Several approaches to the SmBBH and EMRI search problem have been proposed, although none are fully developed.
One family of approaches involves splitting the data into multiple time or frequency segments and searching each individually.
It is not necessarily expected to be possible to confidently detect a signal in a single segment, but by suitably combining the results of searches across segments a detection can be achieved.
This type of method can be described as incoherent, or semi-coherent, because the model used is not required to accurately describe the signal phase evolution across the entire observation \cite{Gair_2004_EMRI_template_bank}.
Semi-coherent methods relax the stringent requirements on the phase accuracy of the models; therefore, another attractive aspect of semi-coherent methods is the prospect of being able to use a simpler, computationally cheaper waveform (e.g.\ a lower-dimensional model, perhaps neglecting some of the physics) for the search. 
Semi-coherent methods are already used in searches for continuous GWs in LIGO/Virgo data \cite{2022_Riles_CW}.
Another approach, specific to EMRIs, is harmonic matching, where several discrete frequency harmonics of the signal are first identified individually before being later combined into a single detection (see Fig.~5.8 of Ref.~\cite{Burke_2021} and Ref.~\cite{Gair_2008_TF_analysis}). 
In practice, this is challenging as individual harmonics are quieter than the full signal and are therefore harder to disentangle from instrumental noise and the numerous other overlapping sources.
In principle, semi-coherent and harmonic matching techniques can be used in combination.
We also note the existence of machine-learning based approaches to the search problem; Ref.~\cite{Zhang_2022_EMRI_detection} demonstrated the detection of high signal-to-noise ratio (SNR) EMRIs using convolutional neural networks, but without the ability to provide information on the source parameters. 

This study focuses on the use of a semi-coherent approach, in combination with a particle swarm optimization (PSO) algorithm to make progress towards a realistic search algorithm for SmBBH signals in LISA. 
PSO is a stochastic optimization algorithm (see, e.g.\ Refs.~\cite{Kennedy_1995_PSO,1998_Shi_PSO}), variants of which can be tailored to be well-suited to the identification of multiple, widely separated peaks in the likelihood surface \citep{Parrot_2006_SPSO,Kennedy_2000_SPSO}. 
It is our hope that this property will also make it suitable for EMRI searches (this will be explored in future work). 
PSO methods have previously been used in a LISA context for galactic double white dwarf binaries \cite{Bouffanais_2016_PSO_GB,Zhang_2021_PSO}.
We show that PSO can successfully locate the source parameters for an SmBBH signal when coupled with a hierarchical approach, iteratively exploring semi-coherent likelihoods with a decreasing number of segments. The work done in this study is still in the early prototyping stage for this search strategy; due to this, several simplifying assumption have been made in the analyses in Secs.\ref{SmBBH_test_bench} and \ref{PySO}. 
These include the use of a simulated data-stream without the presence of noise (i.e.\ working with zero-noise injections) and without other confusing sources such as galactic binaries. We also assume the LISA instrumental noise will be stationary and use a constant power spectral density (PSD), neglecting the expected cyclo-stationary and non-stationary components. Our data-stream is also devoid of gaps and glitches, which will exist in the real LISA data-stream.

The semi-coherent methods that are used in this study are introduced in Sec.~\ref{Theory}. 
Sec.~\ref{sec:GW170817} explores the properties of the semi-coherent likelihood by using it to reanalyze the GW170817 BNS event. 
Sec.~\ref{SmBBH_test_bench} explores the properties of the semi-coherent likelihood for SmBBH sources. 
In Sec.~\ref{PySO} we introduce PSO as a search method which is able to locate the source parameters for a SmBBH signal. 
Sec.~\ref{sec:discussion} discusses the further work required to develop this into a full search and possible extensions of this method to explore the extremely multi-modal likelihood surfaces expected from EMRI signals. 
Throughout this paper we work in natural units where $G=c=1$.

\begin{figure*}[t]
    \centering
    \includegraphics[width=\textwidth]{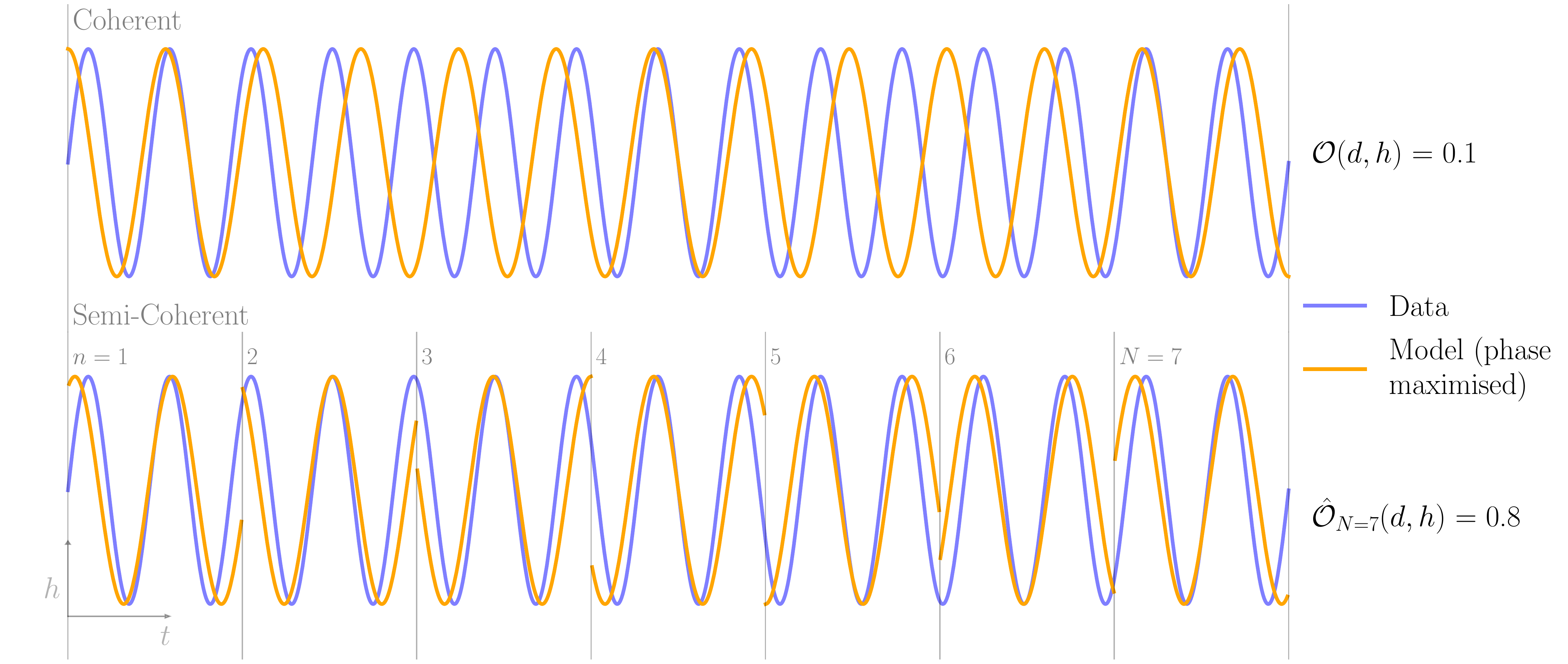}
    \caption{
    A sketch illustrating the semi-coherent method. For simplicity, time-domain sinusoidal signals are used; the data (purple) $d=\sin(2\pi f^{\prime} t)$, where $f^\prime T_{\rm obs} = 15$, and the model (orange) $h=\sin(2\pi ft)$, where $(f-f^\prime)T_{\rm obs}=2.5$. \emph{Top panel:} a coherent analysis; the model is compared against the data, trying to coherently maximize the overlap $\mathcal{O}(d,h)$ across the entire observation period, $T_{\rm obs}$. \emph{Bottom panel:} an $N\!=\!7$ segment semi-coherent analysis; the model phase is varied in each segment (shown in gray) independently to maximize the overlap, leading to discontinuities at the segment boundaries. In the coherent analysis, the model drifts out of phase with the data leading to a low overlap. The extra freedom in the semi-coherent analysis partially compensates for this leading to a larger semi-coherent overlap, $\hat{\mathcal{O}}_{N=7}(d,h)\!\gg\!\mathcal{O}(d,h)$. The amplitudes are normalized such that $\bra{d}\ket{d} = 1$.}
    \label{fig:Semi_coherence_intuition}
\end{figure*}

\section{Semi-Coherent methods}\label{Theory}

In this section we describe the semi-coherent data analysis methods used in this study and contrast them with the conventional, fully-coherent analysis more commonly used in GW astronomy. 

In GW data analysis, the noise-weighted inner product plays a key role in both the search and parameter estimation phases. 
The noise-weighted inner product between two sets of time series $a_\alpha$ and $b_\alpha$ is usually defined in the frequency domain as 
\begin{equation}\label{Noise_weighted_inner_product_eqn}
    \bra{a}\ket{b} = \sum_{\alpha}4 \mathrm{Re} \bigg[\int_{f_{\mathrm{min}}}^{f_{\mathrm{max}}} \frac{a_{\alpha}(f)b^\dagger_{\alpha}(f)}{S_{\alpha}(f)}\mathrm{d}f\bigg],
\end{equation}
where the dagger denotes complex conjugation and $\alpha$ labels different data streams which are assumed to contain independent Gaussian noise with (one-sided) power spectral densities (PSD) $S_\alpha(f)$.
For a network of ground-based detectors these data streams are the measurements from different detectors (e.g.\ $\alpha\in\{\mathrm{H},\mathrm{L},\mathrm{V}\}$). Whereas in LISA they are the noise-orthogonal time-delay interferometry (TDI) channels (e.g.\ $\alpha\in \{\mathrm{X}, \mathrm{Y}, \mathrm{Z}\}$, or $\alpha\in \{\mathrm{A}, \mathrm{E}, \mathrm{T}\}$) which are constructed on the ground from the raw LISA L0 data to suppress the dominant laser noise \cite{2002_Prince_TDI}.

The measured data, $d_\alpha=h_\alpha(\theta)+n_\alpha$, contains signal and noise.
The waveform model describes the signal using the parameter vector $\theta$.
Parameter estimation uses the following log-likelihood which is explored as a function of the model parameters;
\begin{align}\label{eq:Standard_likelihood_eqn}
    \log L(d|\theta) = & -\frac{1}{2}\bra{d-h}\ket{d-h} + c \\
    = & -\frac{1}{2}\bra{d}\ket{d} -\frac{1}{2}\bra{h}\ket{h} + \bra{d}\ket{h} + c . \nonumber
\end{align}
The number of free parameters is the dimensionality of the parameter vector, $\mathrm{dim}(\theta)$.
The log-normalization $c$ does not depend on $\theta$.
On the second line, $\log L$ is split into three terms: $\bra{d}\ket{d}$ is constant (in that is doesn't depend on $\theta$) and can be neglected; $\bra{h}\ket{h}=\rho^2$ is the optimal squared signal-to-noise ratio (SNR) and is approximately constant over small regions of parameter space.
Therefore, $\bra{h}\ket{d}$ is the key quantity that controls the shape of the likelihood surface.

For most calculations in this paper it will be assumed that the signal contains a single mode, by which we mean the model can be decomposed into amplitude and phase as $h(f)=A(f)e^{i\Phi(f)}$ where one of the waveform parameters is an orbital phase angle $\phi$ which enters as $\Phi(f)\rightarrow\Phi(f)+\phi$ (the angle $\phi$ is one component of the parameter vector $\theta$). If this is the case, then it is possible to analytically maximize the $\bra{h}\ket{d}$ term with respect to $\phi$. 
We define the overlap as this phase-maximized inner product;
\begin{align}\label{eq:phase_maximized_inner_product}
    \mathcal{O}(d,h) \equiv & \, \underset{\phi}{\text{max}}\bra{d}\ket{he^{i\phi}}\nonumber \\
    = &4 \bigg|  \sum_{\alpha}  \int_{f_{\mathrm{min}}}^{f_{\mathrm{max}}} \frac{d_{\alpha}(f)h_\alpha^\dagger(f)}{S_{\alpha}(f)}\mathrm{d}f\bigg| .
\end{align}
Note, the overlap is simply the magnitude of a complex inner product.
If the model contains multiple modes, the maximization with respect to $\phi$ must be done numerically.
The coherent overlap in Eq.~\ref{eq:phase_maximized_inner_product} is a function of $\mathrm{dim}(\theta)-1$ free parameters, not including the phase angle $\phi$.

While maximizing over the phase angle will affect the shape of the likelihood (and hence the posterior distribution) in the other parameters and is less desirable than marginalizing over it, the difference is expected to be small, especially for long signals such as SmBBHs and EMRIs. This is demonstrated explicitly in Sec.~\ref{sec:GW170817} for the BNS signal GW170817 and is also found to be the case in Sec.~\ref{SmBBH_test_bench} for SmBBH signals.

We now proceed to split the inner product into $N$ segments.
The segmented inner product between two sets of time series $a_\alpha$ and $b_\alpha$ is calculated in terms of the $N$ separate frequency integrals,
\begin{eqnarray}
    [a|b]^{N}_n = \sum_{\alpha} 4 \mathrm{Re} \int_{f_{n}}^{f_{n+1}} \frac{a_{\alpha}(f)b^\dagger_{\alpha}(f)}{S_{\alpha}(f)}\mathrm{d}f.
\end{eqnarray}
We emphasize that we are segmenting our data in the frequency domain; each segment involves data taken at all times. 
This is to be contrasted with what was envisaged in, for example, Fig.~1 of Ref.~\cite{Gair_2004_EMRI_template_bank}, where the data was segmented in time. For slowly inspiraling sources such as SmBBHs which are well approximated by a stationary phase approximation the two approaches are equivalent.
Here the $f_0 = f_{\mathrm{low}}$, $f_{N} = f_{\mathrm{high}}$, and the intermediate frequency boundaries $f_n$ are ordered 
as $f_n<f_{n+1}$. 
For now the frequency boundaries are only required to be ordered, and we will discuss a method for selecting these later in this section.
Since there is no phase maximization incorporated into this inner product yet, the sum across all segments is equal to the standard noise-weighted inner product,
\begin{eqnarray}
    \sum_{n=0}^{N-1} [a|b]^N_n = \bra{a}\ket{b} .
\end{eqnarray}

So far we have not actually done anything except (arbitrarily) splitting the integral in Eq.~\ref{Noise_weighted_inner_product_eqn} into a number of sub integrals. However, we now generalize by allowing the waveform model to be different in each segment.
The most extreme approach is to allow all of the waveform model parameters to differ in every segment; in the frequency range $f_n<f<f_{n+1}$, the waveform model is $h(\theta_n)$. In this case the total number of model parameters is now $N\mathrm{dim}(\theta)$. Note that the waveform model is discontinuous at the segment boundaries.
Combining the phase-maximized and segmented inner products, we define a semi-coherent overlap with $N$ segments. The phase-maximized inner products for each segment follows the same form as Eq.~\ref{eq:phase_maximized_inner_product} truncated at the appropriate frequency boundaries for that segment:
\begin{eqnarray}\label{eq:semi-coherent_O}
    \hat{\mathcal{O}}_N(d,h) = \sum_{n=0}^{N-1} \underset{\phi_n}{\text{max}}\,[d|he^{i\phi_n}]^N_n
\end{eqnarray} 
Therefore, there are now $N$ phase parameters in our model, one per segment. We maximize over all of these phase parameters independently, each of which individually are unphysical, a combination of which correspond to the orbital phase $\phi$. 
The semi-coherent overlap in Eq.~\ref{eq:semi-coherent_O} is a function of the $N(\mathrm{dim}(\theta)-1)$ free parameters $\{\theta_0,\theta_1,\ldots,\theta_{N-1}\}$, not including the phase angles $\phi_n$.

The purpose of introducing all the additional parameters $\theta_n$ is to make the model less sensitive to any of the parameters individually. For example, in a coherent analysis a small change in, say, the chirp mass parameter, $\mathcal{M}_c$, may be enough to alter the phase evolution of the signal and cause the coherent overlap to drop significantly, $\mathcal{O}\sim 0$, but the same small change in the chirp mass in just the first segment, $\mathcal{M}_{c,0}$, will have a much smaller effect on the semi-coherent overlap, $\hat{\mathcal{O}}_N\sim 1$. It is intended that this drop in required precision will make it easier to perform the initial search. It may also have the benefit of allowing for the use of less accurate waveform models. 

These benefits come at a cost; the total number of free parameters is increased.
This increase in the flexibility of the model necessarily leads to a decrease in sensitivity. 
This flexible model is more likely to be able to fit well a signal containing only noise. 
For this reason, a search algorithm using the semi-coherent likelihood has an increased false alarm rate (compared to a similar search using the normal, coherent likelihood) and must therefore raise the detection threshold accordingly \citep{2017_Chua_Semi_Coherent_detprob,Gair_2004_EMRI_template_bank}. 
In the context of continuous GW searches with ground-based interferometers, it is known that in the limit of a large number of segments ($N\rightarrow\infty$) the sensitivity of an idealized semi-coherent search loses sensitivity and is lower than that of an idealized coherent search by a factor $\propto N^{\frac{1}{4}}$ \cite{2021_Woan_Semi_Coherent}. 
One possible approach would be to lower the threshold sensitivity in the early stages of the search, accepting a larger number of false positives, which are then followed up and can be vetoed in later stages of the search. We leave this for future work.

Note that in the case of $N=1$ segments, the semi-coherent overlap recovers the standard, coherent phase-maximized result;
\begin{align}
    \hat{\mathcal{O}}_{N=1}(a,b)=\mathcal{O}(a,b).
\end{align}

Swapping the semi-coherent overlap into the expression for the log-likelihood in Eq.~\ref{eq:Standard_likelihood_eqn}, we reach our definition of the semi-coherent log-likelihood;
\begin{align} \label{eq:def_semico_L}
    \mathrm{log}\hat{L}_N(d|\theta_0,\theta_1,\ldots,\theta_{N-1}) = &-\frac{1}{2}\bra{d}\ket{d} -\frac{1}{2}\bra{h}\ket{h} \nonumber \\ &+ \hat{\mathcal{O}}_N(d,h).
\end{align}
Note $\bra{h}\ket{h}$ remains a standard inner product because (for a single mode waveform) $\bra{h}\ket{h}$ does not depend on $\phi$. This will not be exactly true for a signal that contains many modes, such as an EMRI.

What has been described so far is an extreme approach to a semi-coherent analysis where all the model parameters are allowed to vary between segments.
This may not be necessary; it is typically the phase evolution that is most important as the overlap is most sensitive to this.
For a signal with many frequency modes such as an EMRI, one might imagine a search keeping the majority of the parameters (e.g.\ sky position, distance, BH masses and spin, and the orbital shape parameters $p$, $e$ and $\iota$; see, for example, Ref.~\cite{2006PhRvD..73b4027D}) constant between segments, while introducing extra phase angles in each segment that can be maximized over.
As we are concerned here with single-mode waveforms, for the remainder of this paper we restrict to the case where only a single orbital phase parameter $\phi_n$ is allowed to vary between segments, and these are maximized over analytically as in Eq.~\ref{eq:phase_maximized_inner_product}. Therefore, our semi-coherent likelihood becomes
\begin{align} \label{eq:def_semico_L_2}
    \mathrm{log}\hat{L}_N(d|\theta) = &-\frac{1}{2}\bra{d}\ket{d} -\frac{1}{2}\bra{h}\ket{h}  + \hat{\mathcal{O}}_N(d,h),
\end{align}
which is a function of just $\mathrm{dim}(\theta)-1$ parameters.
Note, that in the case of $N=1$ segments, the semi-coherent likelihood is related to the standard, coherent likelihood via
\begin{align}
    \mathrm{log}\hat{L}_{N=1}(d|\theta) = \underset{\phi}{\text{max}} \, \mathrm{log}L(d|\theta) .
\end{align}

Eq.~\ref{eq:def_semico_L_2} is the definition of $\hat{L}$. We name this quantity the \emph{semi-coherent likelihood} emphasizing the connection with $L$ in Eq.~\ref{eq:Standard_likelihood_eqn}.
However, $\hat{L}$ is not a likelihood in the usual sense. Because it is a function of the data it can be regarded as a new statistic that is introduced here as part of a new proposed search strategy.

Fig.~\ref{fig:Semi_coherence_intuition} illustrates the semi-coherent approach for a pair of simple sinusoidal waves with similar, but not identical frequencies. In the figure the idea is illustrated in the time domain, although our analysis in the following sections will segment the data in the frequency domain.
The signal and data gradually drift out of phase with each other over many cycles resulting in a low coherent overlap.
In the semi-coherent analysis the model frequency is kept constant across the entire range of the observation but the phase angle is allowed to vary between segments;
this partially compensates for the difference in frequency with the data and the semi-coherent overlap is much higher than the vanilla inner product. 
The semi-coherent overlap is less sensitive to variation of the parameters that affect the frequency and phase evolution of the signals. 

There is a freedom in our definition of the semi-coherent likelihood corresponding to the choice of the segment boundaries $f_n$.
Perhaps the simplest option is uniform segmentation with $f_n=f_{\rm min}+(n/N)(f_{\rm max}-f_{\rm min})$. 
However, SmBBH and EMRI systems spend a disproportionately large amount of time at lower frequencies. Therefore, uniform segmentation would result in the majority of the signal being contained in a small number of segments. Another option is logarithmic segmentation with $\log (f_n/f_{\rm min}) = (n/N)\log(f_{\rm max}/f_{\rm min})$.
This results in more segments at lower frequencies, however the signal is still not necessarily split equally between the segments.
Therefore, we choose to define our segment boundaries with respect to the signal that is being analyzed.
We opt to define segments such that they contain an equal squared SNR.  
The squared SNR in segment $n$ is 
\begin{eqnarray}\label{eq:equal_sq_snr}
    \rho^2_n \equiv \sum_\alpha 4 \int_{f_n}^{f_{n+1}}\frac{|h_\alpha(f)|^2}{S_\alpha(f)}\mathrm{d}f = \frac{\rho^2}{N};
\end{eqnarray}
this is an implicit equation for the segment boundaries $f_n$ under the equal squares SNR segmentation scheme.
The total (optimum) SNR is given by $\rho^2=\sum_n\rho^2_n$. 
The downside of this approach is that the segment boundaries $f_n(\theta)$ now depend on the source parameters, and must be recomputed at each evaluation of the semi-coherent likelihood.

\begin{figure*}[t!]
    \centering
    \includegraphics[width=\textwidth]{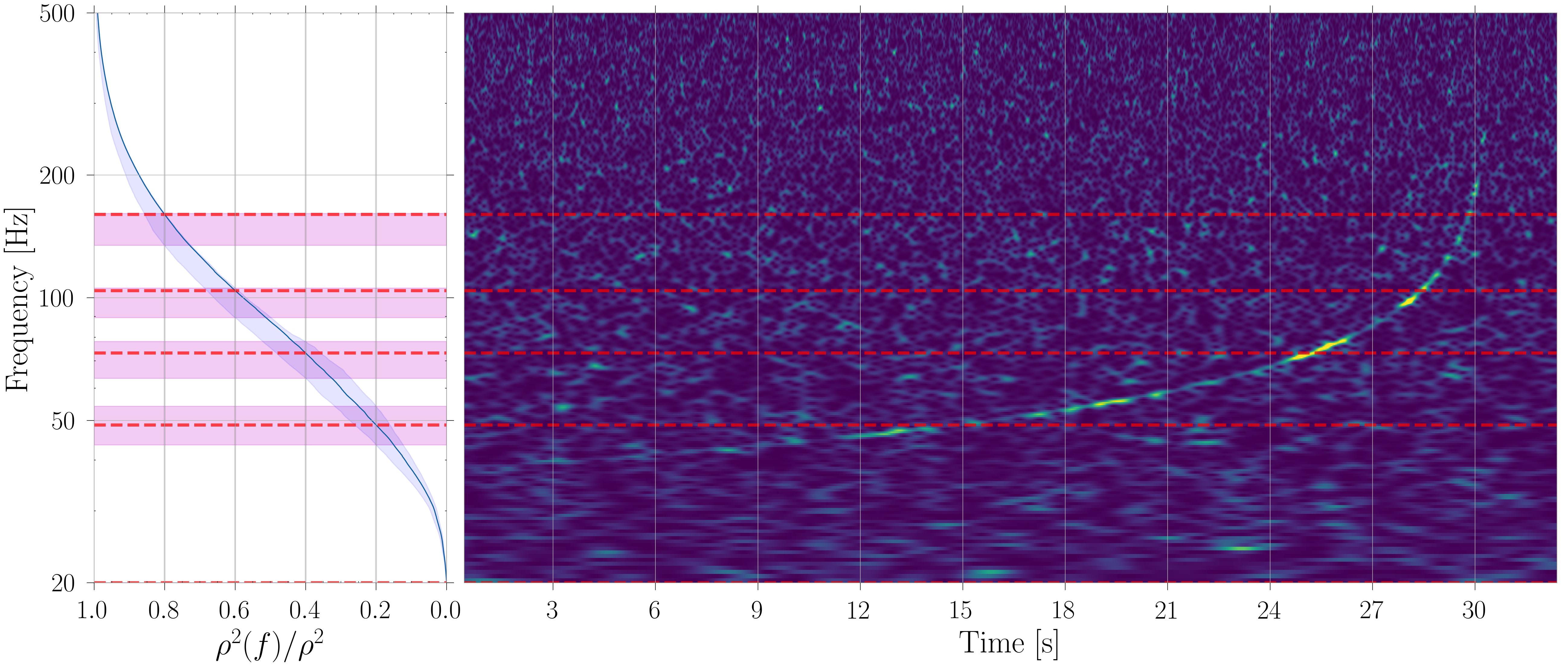}
    \caption{
    Illustration of the $N=5$ segment semi-coherent analysis of GW170817.
    \emph{Right panel:} The $Q$-transform time-frequency scan of the $32\,\mathrm{s}$ of analysis data from the LIGO Livingston detector (other two instruments not shown). The characteristic chirp of the BNS is clearly visible.
    \emph{Left panel:} The cumulative (normalized) squared network SNR. The rate at which the SNR accumulates with frequency depends on the source parameters; the blue line and shaded region show the result for the maximum likelihood and 90\% credible region computed using the posterior samples from Sec.~\ref{sec:GW170817}.
    The $N=5$ segment semi-coherent likelihood splits the frequency range $[20,1000]\,\mathrm{Hz}$ at 4 intermediate frequencies such that an equal squared SNR accumulates in each segment.
    These four frequencies (median and 90\% credible regions) are shown with horizontal red lines. 
    } 
    \label{fig:Q_transform_BNS}
\end{figure*}

It is hoped that the semi-coherent likelihood will lead to wider posterior distributions, particularly on those parameters which strongly influence the phase of the GW signal (these are referred to as \emph{phasing parameters}). For the signals observed by ground-based detectors the phasing parameters can be identified with the \emph{intrinsic} source parameters. However, this identification breaks down for LISA where extrinsic parameters (such as the sky position) also effect the phase (e.g.\ via the direction-dependent Doppler shift due to the motion of the detector). 
It is shown in Sections ~\ref{sec:GW170817} and ~\ref{SmBBH_test_bench} that the semi-coherent analysis method does indeed broaden the posteriors on the most important phasing parameters while leaving the posteriors on the other parameters largely unchanged.

Another well-known method for broadening posterior distributions is tempering or simulated annealing \cite{Swendsen_1986_PT,Earl_2005_PT}; this involves raising the likelihood to a power $\beta$, where $\frac{1}{\beta}$ is commonly called the annealing temperature. 
This can be used as a method of accelerating sampling for highly multimodal probability distributions, because it makes it easier for many stochastic methods to traverse the likelihood surface. This has been used in several parameter estimation studies for sources in LISA data \cite{Tyson_2024_Global_Fit,Digman_2022_SmBBH,Sberna_2022_190521}.
Tempering or annealing modifies the likelihood surface in a way that is somewhat similar to the semi-coherent approach, in that it reduces the severity of secondary maxima.
However, there exists a clear distinction between the two methods. The semi-coherent method smooths the log-likelihood surface around the injection, \emph{removing} secondary peaks in the log-likelihood surface, whereas tempering raises the ``floor'' value of the log-likelihood at large distances from the peak which has the effect of gradually congealing secondary peaks in the likelihood. 
Tempering preserves multi-modality in the log-likelihood surface while the semi-coherent method eradicates it.
See appendix \ref{tempering_comparison}, and Fig.~\ref{fig:Semi_coherent_vs_tempering} therein for a comparison between tempering and semi-coherent methods.

\section{Case study: the binary neutron star GW170817}\label{sec:GW170817}

\begin{figure*}[t!]
    \centering
    \subfloat{\includegraphics[width=0.41\textwidth]{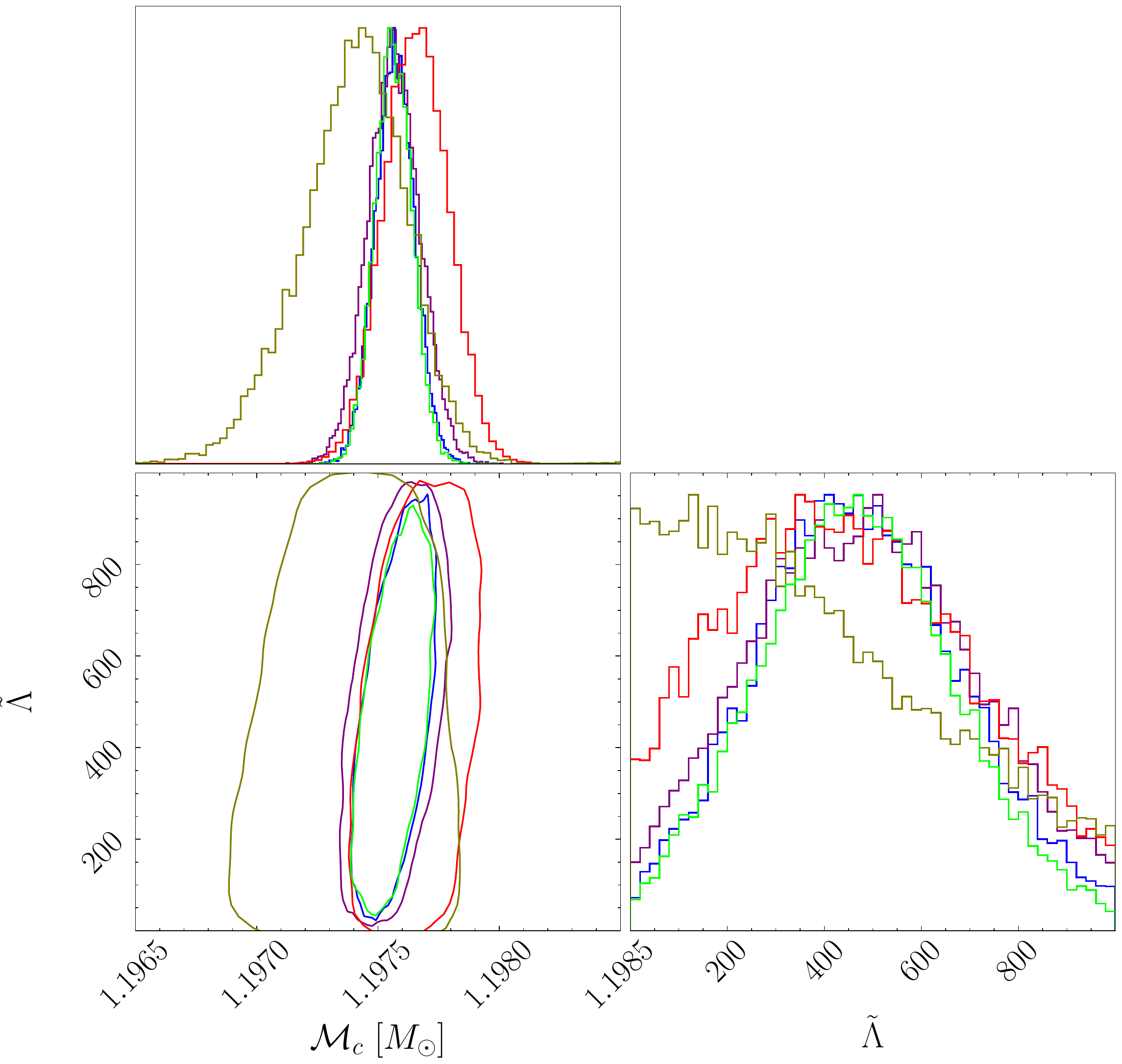}}
    \hspace{1cm}
    \subfloat{\includegraphics[width=0.5\textwidth]{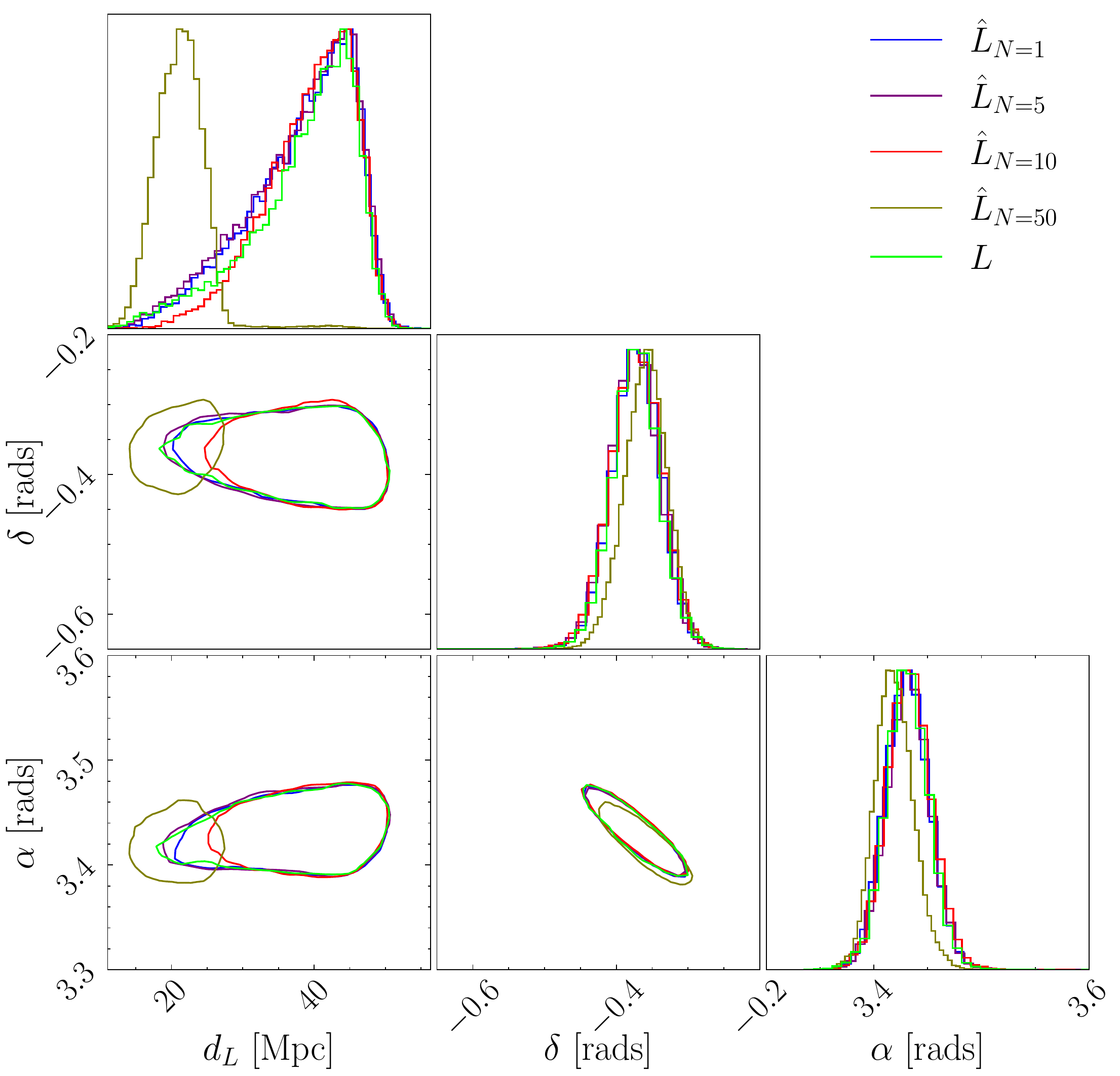}}
    \caption{Posterior distributions obtained with various semi-coherent likelihoods $\hat{L}_N(d|\theta)$ with $N$ in the range 1 to 50, and the vanilla likelihood $L(d|\theta)$. \emph{Left panel:} posterior distributions for a selection of the phasing parameters ($\mathcal{M}_c$ and $\Tilde{\Lambda}$). These are intrinsic parameters describing the frequency (or phase) evolution of the source. \emph{Right panel:} posterior distributions on a selection of non-phasing parameters (luminosity distance $d_L$, declination $\delta$, and right ascension, $\alpha$). These are extrinsic parameters describing the location of the source. 
    In all 2D plots the contours show the 90\% confidence regions. 
    }
    \label{fig:BNS_posteriors}
\end{figure*}

The BNS signal GW170817 is the longest GW signal observed to date. In some respects it is the closest thing we currently have to a SmBBH or EMRI signal (albeit, still with orders of magnitude fewer wave cycles). Therefore, reanalyzing GW170817 is a gentle way to prepare and build up to analyzing SmBBH and EMRI signals. In this section we explore the properties of the semi-coherent likelihood in Eq.~\ref{eq:def_semico_L_2} by using it to reanalyze GW170817.

This case study is intended to build intuition for the semi-coherent likelihood, ensuring that the peaks, although broader, remain consistent with the vanilla likelihood (i.e.\ $L$ in Eq.~\ref{eq:Standard_likelihood_eqn}). 
This also gives us an opportunity to explore how the semi-coherent likelihood behaves in the presence of real detector noise. Although a real search will aim to just locate the peaks in the semi-coherent likelihood, in this section the full likelihood distribution is explored using stochastic sampling, in order to gain a better understanding of the tails of the semi-coherent likelihood surface.

In the following analysis we use the waveform model \texttt{IMRPhenomPv2$\_$NRTidalv2} \cite{2019_Dietrich_NRTidalV2, 2019_Dietrich_NRTidal}. 
This is a fast, frequency domain, phenomenological waveform model built on the quasi-circular, spin-precessing binary BH model \texttt{IMRPhenomPv2} \citep{2012_Schmidt_PhenomPv2,2014_Hannam_PhenomPv2}. Tidal effects are expected to be significant in the late inspiral of a BNS system and \texttt{IMRPhenomPv2$\_$NRTidalv2} accounts for this through the inclusion of tidal deformability parameters for both compact objects; these are parametrized by $\tilde{\Lambda}$ and $\delta \tilde{\Lambda}$ \cite{Chatziioannou_2020_BNS_tidal_phasing}. 
The \texttt{PhenomPv2} waveform model is in turn constructed from the spin-aligned \texttt{PhenomD} \citep{Husa_2016_PhenomD,Khan_2016_PhenomD} model which contains only the $(l,|m|)=(2,2)$ mode. \texttt{PhenomPv2} ``twists'' this model in a way that mimics the effects of spin-orbit precession. 
It is the $(2,2)$ mode of the \texttt{PhenomD} waveform that the \texttt{NRTidalv2} model modifies, incorporating an amplitude and phase correction which originate from the tidal interactions between the two neutron stars in the binary.

The data for the following analysis span $32$ seconds in the GPS time range $[1187008852.4, 1187008884.4]\rm{s}$. The PSD is estimated using a Welch periodogram (as implemented in GWpy \cite{Macleod_2021_GWpy}) using $1024$ seconds of off-source data in the GPS time range $[1187007316.4, 1187008340.4]\rm{s}$, which is offset from the trigger by $512\,\mathrm{s}$ to avoid any possible contamination from the long-lived signal. 
The time-series data, sampled at $4096\,\mathrm{Hz}$, was obtained from the Gravitational Wave Open Science Center \cite{2021_GWOSC}.  
Data from the Livingston, Hanford and Virgo detectors are used.
The Livingston data contains a prominent glitch just before the merger (see Fig.~2 of Ref.~\cite{Abbot_2017_BNS}). The glitch has been modeled and removed using \texttt{Bayeswave} \cite{2015_Cornish_Bayeswave}. Specifically the glitch-subtracted data is obtained from Ref.~\cite{GW170817_data}. The data was analyzed using $f_{\rm min}=20\,\mathrm{Hz}$ to $f_{\rm max}=1000\,\mathrm{Hz}$.

The likelihoods used in this case study segment the data and model using the equal square SNR scheme discussed in Sec.~\ref{Theory}. 
At each new evaluation of the likelihood, i.e.\ at each proposed set of parameters $\theta$, the segment boundaries $f_n$ must be recomputed from Eq.~\ref{eq:equal_sq_snr}. 
This is done using the cumulative squared SNR as a function of frequency;
\begin{eqnarray}
    \rho^2(f) = \sum_{\alpha} 4\int_{f_{\rm min}}^f \frac{|h_{\alpha}(f)|^2}{S_{\alpha}(f)} \mathrm{d}{f},
\end{eqnarray}
where $\rho^2\equiv\rho^2(f_{\rm max})$. This curve is used to divide the total square SNR into segments that contain equal $\rho^2_n$. This process is illustrated in Fig.~\ref{fig:Q_transform_BNS}.

Likelihood maximization with respect to phase is performed analytically, as shown in Eq.~\ref{eq:phase_maximized_inner_product}. 
This analytic maximization is possible because the \texttt{PhenomD} model, from which our waveform is constructed, is a single mode $(2,2)$ waveform. 
It has been verified that the likelihood is unchanged if the phase maximization is instead performed numerically. 

Stochastic sampling of the posterior distribution was performed using the \texttt{dynesty} \cite{Speagle_2020_dynesty} nested sampler \cite{2006_Skilling_nest_sampling} as implemented in the \texttt{Bilby} \cite{Ashton_2019_bilby} library.
However, we note that a custom log-likelihood function is used, which implements the semi-coherent $\log\hat{L}_N$ described in Sec.~\ref{Theory}.
The priors used for the following analyses were those in Ref.~\cite{GW170817_Bilby_analysis}, with the exceptions of the following parameters: the priors on the sky position angles (right ascension $\alpha$ and declination $\delta$)  are uniform over the whole sky ($\cos\delta \in [-1,1]$ and $\alpha \in [0,2\pi]$), the dimensionless tidal deformability parameters are uniform over the range $\tilde{\Lambda}\in[0,1000]$ and $\delta\tilde{\Lambda}\in [-5000,5000]$.

Fig.~\ref{fig:BNS_posteriors} shows a set of posterior distributions obtained with both the semi-coherent and vanilla likelihoods. The corner plots show subsets of the source parameters, comprised of phasing (left) and non-phasing (right) parameters. Two important phasing parameters in this case are the chirp mass ($\mathcal{M}_c$) and the dimensionless tidal deformability parameter ($\Tilde{\Lambda}$). Examples of non-phasing parameters are those that define the $3$D location of the source. Posterior distributions on the phasing parameters broaden as the number of segments used in the semi-coherent likelihood is increased, while the distributions for the non-phasing parameters are not significantly affected. The exception to this is the luminosity distance posterior for the $N=50$ segment likelihood, which is biased, and is not consistent with lower segment posteriors (which are themselves consistent with literature \citep{Abbot_2017_BNS,GW170817_Bilby_analysis}). We have verified that the distance posterior varies continuously between the $N=10$ and $N=50$ cases shown. The bias in the distance will not be problematic for the purpose of a search.
Fig.~\ref{fig:BNS_posteriors} displays several posteriors obtained from the same underlying data, which exhibits identical noise characteristics, however the likelihoods are different; the small shifts between the likelihood peaks can likely be attributed to the semi-coherent nature of each likelihood interacting with the noise.  

The semi-coherent likelihood partitions the data into a number of segments. A natural theoretical maximum number of such segments is set by the number of orbits in the signal. Beyond this number, each frequency segment covers $\sim$ one orbit and orbital phase maximization ceases to be meaningful. A practical, useful maximum number of segments will be a correction factor  $(\gamma<1)$ multiplied by the number of orbits in the signal. In the case of GW170817, the signal has $\sim 3000$ orbits in band over the whole $\sim 100$ seconds the signal is present in the detector \cite{Abbot_2017_BNS}. 
The semi-coherent limit for the number of segments has been verified; we checked that the semi-coherent analysis with $\sim 500$ segments produces posteriors that are extremely broad, multi-modal, and not consistent with the literature (these results are not shown here), for GW170817 $\gamma \sim 1/6$. SmBBH/EMRI systems have $\gtrsim 10^5$ orbits in band, so the natural practical upper limit of segments will be much higher.

The key takeaway from this case study is the semi-coherent likelihood does broaden the posterior distributions of parameters that control the GW phase, while not affecting the non-phasing parameters. Additionally, while this analysis method breaks down when the number of segments approaches the number of orbital cycles, we hope this will not be an issue for the SmBBH/EMRI signals as both source types will undergo a much larger number of orbits, $\sim 10^5$, within the LISA frequency band.

\section{Stellar-mass binary black holes in LISA}\label{SmBBH_test_bench}

We now apply semi-coherent likelihoods to the analysis of a LISA SmBBH signal. In contrast to the analyses in Sec.~\ref{sec:GW170817} which worked with real noisy data, all of the semi-coherent analyses in this section and Sec.~\ref{PySO} are performed on zero-noise mock injections; this means that the (fully-coherent) likelihood surface is peaked at the injected source parameters. As we are in the early prototyping phase of testing this search strategy, we simulate a noise-less data-stream to simplify the search. We do not expect the inclusion of noise to change the results significantly, posteriors obtained from noisy data will be shifted relative to the true parameters, however the shifted posterior will still be consistent with the no-noise posterior.

The analysis in this section uses parameter estimation methods that fully explore the likelihood distributions (including their low-probability tails) mirroring the analysis performed in the previous section for GW170817. 
This is done to build an understanding of the properties of the semi-coherent likelihood.
The sampling iterates through a sequence of semi-coherent likelihoods with steadily decreasing number of segments, $N$, progressively localizing the signal to smaller regions in parameter space, thereby mimicking a search process. 
However, a real search would not use sampling methods that waste time exploring the tails of the distributions at early stages.
Sec.~\ref{PySO} repeats this procedure using an optimizer (as opposed to a sampler) as part of a more realistic search algorithm.

We inject a fiducial SmBBH source to test our sampling and optimization methods.
The injected source parameters are given in Tab.~\ref{tab:Injection}. 
The sampling is performed over the following parameters with flat priors: chirp mass $\mathcal{M}_c$, time to merger $t_c$, dimensionless mass difference $\delta\mu$, ecliptic longitude $\lambda$, sine of ecliptic latitude $\sin\beta$, square root of left and right-handed circularly polarized GW amplitudes $A^{1/2}_{\mathrm{left},\mathrm{right}}$, dimensionless aligned spin magnitudes $\chi_1$ and $\chi_2$, and phases for the left- and right-handed GW polarizations $\phi_{\mathrm{left},\mathrm{right}}$. 
These are related to the more familiar component mass parameters $m_1$, $m_2$, phase and polarization angles $\psi$, $\phi$, inclination $\iota$, and luminosity distance $d_L$, via the equations in appendix \ref{app:Param_transforms}.

The wide (i.e.\ uninformative) prior ranges $\Delta\theta\equiv\theta_{\rm max}-\theta_{\rm min}$ are chosen to be representative of a search,
Much narrower priors are typically used in parameter estimation studies, e.g.\ Refs.~\citep{2022_Klein_Multiband,Buscicchio_2021}. Most parameters are allowed to vary over their full physical ranges, the exceptions are the important phasing parameters, $\mathcal{M}_c$ and $t_c$. The priors on these parameters are wide enough to cover a sizable fraction ($\sim 1/50$ in both dimensions) of the LISA discovery space described at the beginning of Sec.~\ref{sec:introduction}. We envision eventually using multiple ($\sim 50^2=2500$) such searches to tile the full parameter space.

\begin{table}[t]
\caption{\label{tab:Injection}
    Injection parameters and priors for both sampling and optimization conducted in sections \ref{SmBBH_test_bench} and \ref{PySO}.
    Parameters above the line are those that are sampled in, using flat priors over the ranges shows, those below the line are \emph{derived} parameters defined in the text. All masses are given as detector frame quantities.
}
\begin{ruledtabular}
\begin{tabular}{ccc}
    \textrm{Parameter} &
    \textrm{Injection} & 
    \textrm{Prior range: $\theta_{\rm min}\textrm{--}\theta_{\rm max}$}\\
    \colrule
    $\mathcal{M}_c\,[M_\odot]$ & $62.46453697$ & $[61.46, 63.46]$\\
    $t_c\,[\mathrm{months}]$ & $38.04$ & $[t_c-1, t_c+1]$\\
    $\delta\mu$ & $0.27$ & $[0,0.7]$\\
    $\lambda\,[\mathrm{rad}]$ & $2.0$ & $[0,2\pi]$\\
    $\sin\beta$ & $0.3$& $[-1,1]$\\
    $\sqrt{A_{\mathrm{left}}}\big[\mathrm{pc}^{-1/2}\big]$ & $3.73\times10^{-5}$ & $[0,10^{-4}]$\\
    $\sqrt{A_{\mathrm{right}}}\big[\mathrm{pc}^{-1/2}\big]$ & $4.44\times10^{-5}$  & $[0,10^{-4}]$\\
    $\chi_{1}$ & $-0.58$ & $[-1,1]$\\
    $\chi_{2}$ & $-0.17$ & $[-1,1]$\\
    $\phi_{\mathrm{left}}\,[\mathrm{rads}]$ & $6.04$ & $[0,2\pi]$\\
    $\phi_{\mathrm{right}}\,[\mathrm{rads}]$ & $2.24$ & $[0,2\pi]$\\
    \hline
    $d_L\,[\mathrm{Mpc}]$ & $300$ & -\\
    $m_1\,[M_\odot]$ & $95$ & -\\
    $m_2\,[M_\odot]$ & $55$ & -\\
    $\phi\,[\mathrm{rads}]$ & $1$ &  - \\
    $\psi\,[\mathrm{rads}]$ & $-2.52$& - \\
    $\iota\,[\mathrm{rads}]$ & $1.66$& - \\     
    $\rho$ & $11.44$& - 
\end{tabular}
\end{ruledtabular}
\end{table}

Computationally efficient post-Newtonian waveforms for the inspiral phase of SmBBH systems, incorporating the effects of eccentricity and spin-precession, are available; see for example Refs.~\citep{2021_Klein_wf,2014_Klein_fast_precessing_inspirals,2004_Damour_Eccentric_SmBBH}. 
These low-order post-Newtonian waveforms are expected to be sufficiently faithful for the analysis of SmBBH sources in LISA \cite{2019_Mangiagli_PN_phase_requirements}. These waveforms are computationally fast which makes their use for searching and sampling over the large parameter space feasible.
As a proof of concept, this analysis uses the simple \texttt{TaylorF2} waveform (as implemented in \texttt{Balrog}\footnote{\texttt{Balrog} is a package being developed for waveform generation and parameter estimation for LISA sources, including supermassive binary BH mergers \cite{Pratten_2022}, double white dwarfs \citep{2019_Buscicchio_DWD,2020_Roebber_DWD_Balrog,2020_Korol_DWD,2022_Finch_DWD} and SmBBH inspirals \citep{Buscicchio_2021,2022_Klein_Multiband}.}) for spin-aligned, non-eccentric (i.e.\ quasicircular) binaries. 
Our methods are expected to generalize easily to waveforms which incorporate additional physics, such as spin-orbit precession and eccentric orbits.
The \texttt{TaylorF2} waveform includes only the $(\ell,|m|)=(2,2)$ spherical harmonic mode; this is expected to be sufficient as higher modes are strongly suppressed early in the inspiral. The waveform model produces the polarizations $\tilde{h}_{+,\times}(f)$ as a function of the source parameters, $\theta$.

It is also necessary to model the response of the LISA instrument to the two GW polarizations $\tilde{h}_{+,\times}(f)$.
The three satellites in the LISA constellation are connected by six laser links. The measured phase time series in these links are expected to contain large-amplitude laser noise.
Therefore, six links are combined into three output channels in a process called time-delay interferometry \cite{2021_Tinto_TDI} that is designed to suppress this laser noise below the level of other, secondary noise sources.
The LISA response is modeled using a rigid adiabatic approximation \cite{2004_Rubbo_rigid_body_approximation} which is used produce the TDI outputs $\tilde{h}_{X,Y,Z}(f)$. The response derived from the rigid adiabatic approximation has previously been tested for SmBBH sources in LISA \cite{Buscicchio_2021}. The TDI outputs are then transformed \cite{2002_Prince_TDI} to the noise-orthogonal TDI channels $\tilde{h}_\alpha(f)$, where $\alpha\in\{A,E,T\}$ which are used for the likelihood evaluation.

\begin{figure}
    \centering
    \includegraphics[width=0.5\textwidth]{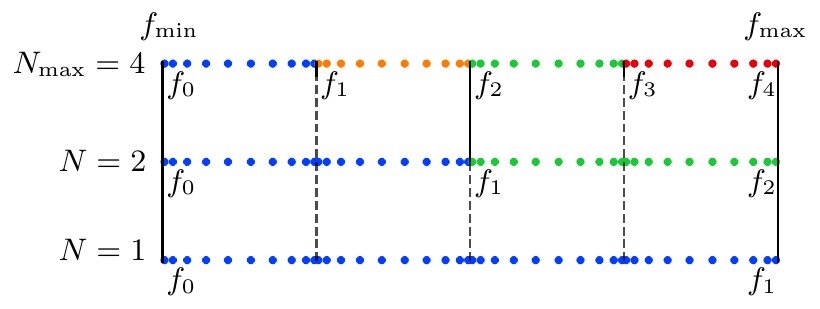}
    \caption{Illustration of the segmentation method used for SmBBH sources with the Clenshaw-Curtis quadrature integration rule. 
    In this illustrative example, the highest number of segments is $N_{\rm max}=2^2=4$, and the other possible numbers of segments ($N=2^1=2$ and the fully coherent $N=2^0=1$) are constructed by the union of pairs of quadrature grid. At each level, colors and solid vertical lines indicate the semi-coherent frequency segments, color dots represent the location of the quadrature integration nodes (in this illustrative example there are 10 nodes per quadrature grid), and dashed vertical lines represent the end of a quadrature integration grid (but not of a semi-coherent segment).
    }
    \label{fig:CCsegmentation}
\end{figure}

\begin{figure*}[t!]
    \centering
    \subfloat{\includegraphics[width=0.45\textwidth]{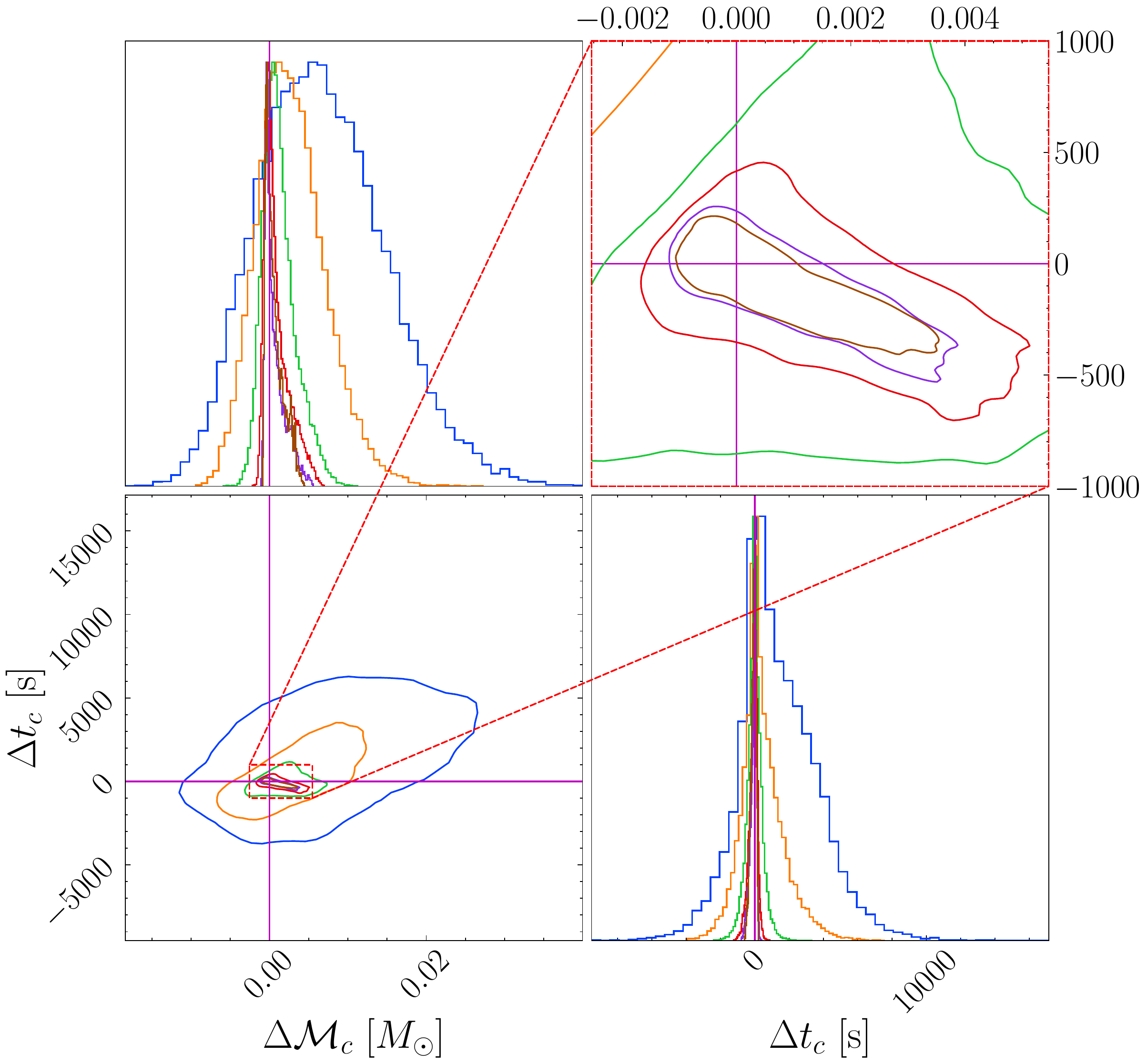}}
    \subfloat{\includegraphics[width=0.45\textwidth]{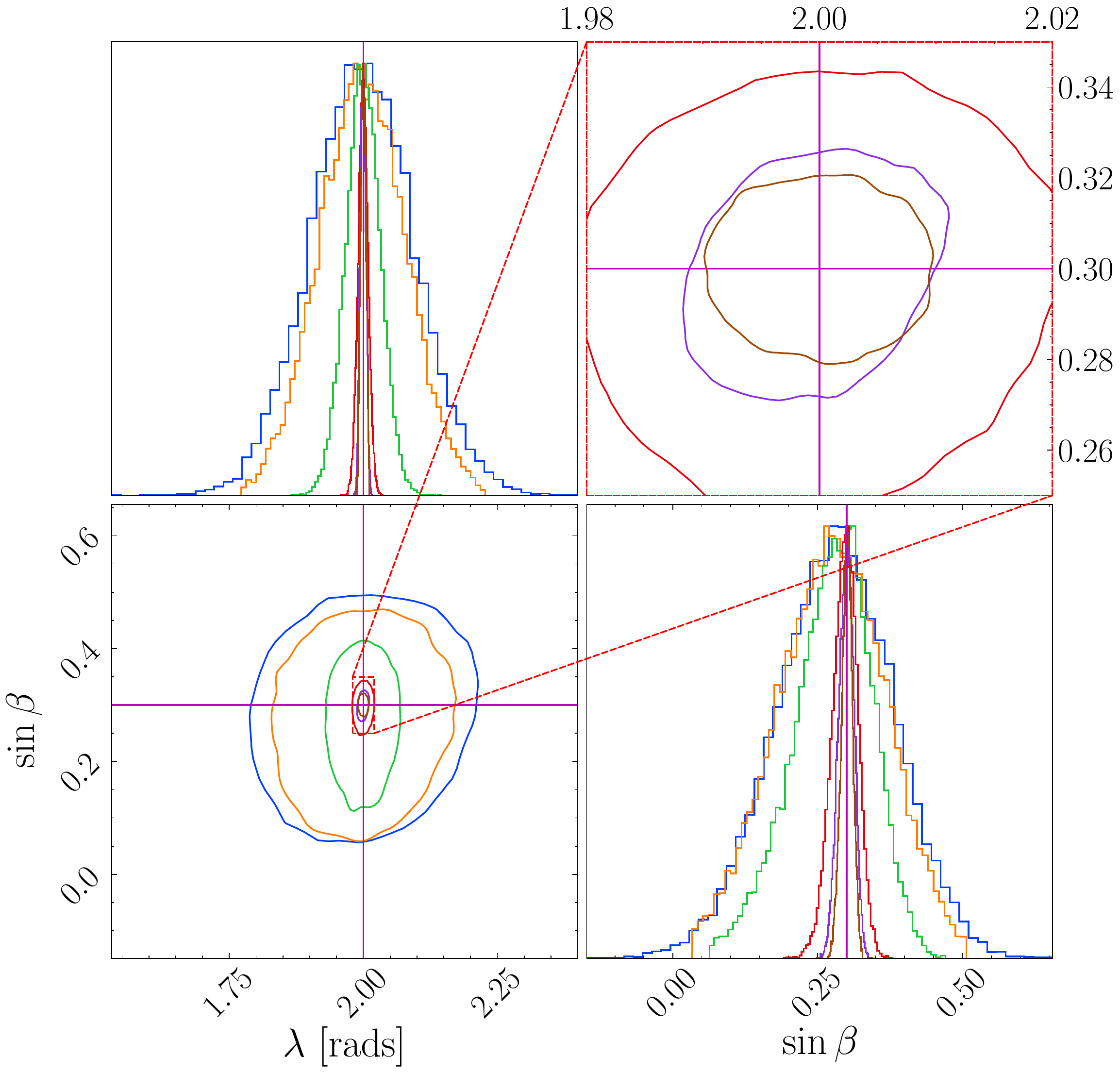}}
    \vspace{-0.6cm}
    \subfloat{\includegraphics[width=0.9\textwidth]{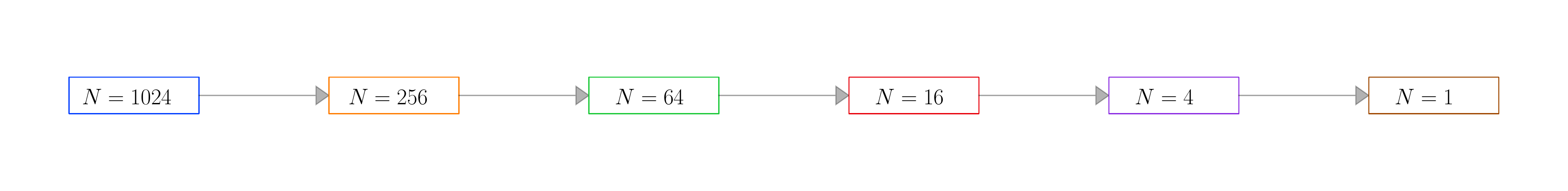}}
    \vspace{-0.5cm}
    \caption{ \label{fig:Inference_hierarchichal}
        Iterative sampling results for the semi-coherent likelihood with varying number of segments, $N$.
        1D marginalized posteriors are shown for chirp mass and time-to-merger parameters (left) and sky position parameters (right). 
        (The axes on the left-hand plot are shifted such that the injection parameters are at the origin.)
        The semi-coherent likelihood with large $N$ broadens the likelihood peak significantly in the parameters which effect the GW phasing; for long-lived LISA sources, such as SmBBHs, this includes the sky position parameters.
        2D contours show the $90\%$ confidence interval. 
        Results are shown for $N\in\{1024,256,64,16,4,1\}$ segments.
        In both corner plots the top-right panel shows a zoom in on the region around the injection parameters.
    }
\end{figure*}

The BNS analysis in Sec.~\ref{sec:GW170817} used a uniformly-sampled fast Fourier transform (FFT) frequency grid. However, SmBBH signals are broadband ($\Delta f\sim 10^{-2}\,\mathrm{Hz}$), and LISA observations are long duration ($T\sim 10^{8}\,\mathrm{s}$), which would result in an FFT frequency grid with $\sim 10^6$ nodes. 
A likelihood calculated using this frequency grid would be too slow to be used in a search.
Instead, \texttt{Balrog} uses Clenshaw-Curtis quadrature \cite{1960_Clenshaw} to accelerate evaluations of the inner product, using $\sim 10^2$ frequency nodes, vastly reducing the computational cost of likelihood evaluations \citep{2022_Klein_Multiband,Buscicchio_2021,Pratten_2022}. We highlight a caveat that quadrature methods are not suitable for evaluating likelihoods on realistic data containing noise, this is further discussed in more detail in appendix \ref{Clensahw_curtis_comments}.

As described in Sec.~\ref{Theory}, we aim to split the data into frequency segments containing equal square SNR.
However, this is complicated by the fact we are no longer integrating using a uniformly-spaced FFT frequency grid. 
To circumvent this, we use multiple quadrature grids adapted to our semi-coherent frequency segments.
Specifically, (i) we choose the maximum number of segments that will be used to be a power of two, $N_{\rm max}=2^a$. (ii) We select a reference waveform with parameters chosen in the center of the prior ranges. The reference waveform is evaluated once (before the search) on the uniformly-sampled FFT grid and this is used to find the segment boundaries $f_0,f_1,\ldots,f_{N_{\rm max}}$ as described in Sec.~\ref{Theory}. 
(iii) We then construct $N_{\rm max}$ irregularly-spaced quadrature frequency grids for these segments.
These frequency grids are then fixed and the data and model (for any parameters in the prior range) are evaluated on this sparse grid. 
For any power-of-two number of segments, $N=2^b$ with $b\leq a$, $\hat{L}_N$, can be evaluated by pairing together these segments. This construction is illustrated in Fig.~\ref{fig:CCsegmentation} with $N_{\rm max}=4$. 
Since, for computational efficiency, the same frequency grid is used for all sources within the prior, most sources have only approximately equal squared SNR per segment. For our fiducial source, prior range and choice of $N_{\rm max}$, we have verified that $\rho_n^2$ varies between segments by a factor $\lesssim 3$.

The semi-coherent likelihood, as defined in Eq.~\ref{eq:def_semico_L_2}, is evaluated using the segmented inner products computed using Clenshaw-Curtis quadrature (see appendix \ref{Clensahw_curtis_comments}),
\begin{eqnarray}
    [a|b]^{N}_n = \sum_{\alpha} 4 \mathrm{Re} \sum_{i}  \frac{w_{i,n} a_{\alpha}(f_{i,n})b^\dagger_{\alpha}(f_{i,n}) }{S_{\alpha}(f_{i,n})}.
\end{eqnarray}
Here, $f_{i,n}$ is the $i^{\rm th}$ node in the the $n^{\rm th}$ frequency quadrature grid and $w_{i,n}$ are the associated quadrature weights. 
If the number of nodes and the span of each quadrature grid is the same, then $w_{i,n}=w_i$. We find empirically that a search starting with $N_{\rm max} = 1024$ segments, with $11$ nodes per quadrature grid, performs well for this source. We note that this is reasonably consistent with the rough early estimate of $N\gtrsim 100$ for the minimum number of segments required for an EMRI search made in Ref.~\cite{Gair_2004_EMRI_template_bank}.

While we choose to use quadrature rules, as illustrated in Fig.~\ref{fig:CCsegmentation}, the semi-coherent method is more general and could be adapted to work with other techniques such as heterodyning/relative binning \cite{Cornish_2021}.

\begin{figure*}[t]
    \centering
    \includegraphics[width=\textwidth]{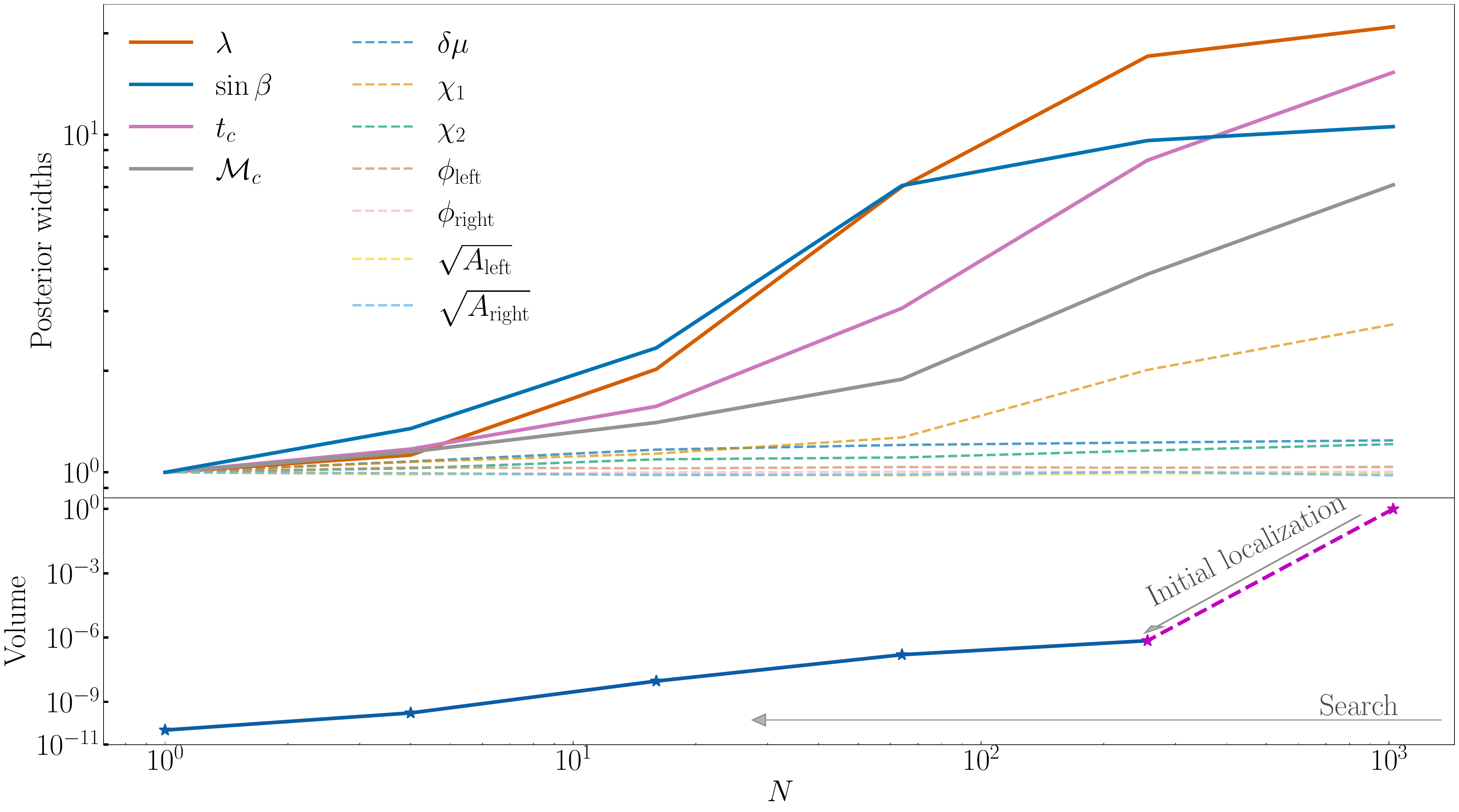}
    \caption{\emph{Top}: The standard deviation, or width, of the 1-dimensional marginalized posterior distributions obtained with the semi-coherent likelihood using different numbers $N$ of segments.
    Posterior widths (normalized by dividing by the width of $N=1$ segment posterior) are shown for all parameters, with the most important phasing parameters highlighted using solid lines. 
    \emph{Bottom}: the volume of the prior hyper-rectangle used for the analysis at each number of segments. Note, only the priors on the four important phasing parameters are changed between each iteration. 
    The volumes are normalized to the initial prior volume at $N=1024$ segments. 
    The initial prior (for $N=1024$) is \emph{not} a measure of the posterior size at this segment, instead it is a prior we have chosen that is large enough to represent the search problem while still remaining tractable for a proof of concept study. We label the transition: $N:1024\rightarrow256$ as the `initial localization' and distinguish it from the other segment transitions using a dashed magenta line.} 
    \label{fig:Inference_posterior_widths}
\end{figure*}

We simulate a $4$ year LISA mission. Inner products are evaluated between the frequency limits $[f_{\text{low}},f_{\text{high}}] = [0.0056,0.1]\,\rm{Hz}$. The source in Tab.~\ref{tab:Injection} is $3.17$ years from merger when LISA observations begin, and is initially radiating at a frequency above $f_{\text{low}}$. After exiting the LISA frequency band, the source merges in $\sim 3.5$ days.
A simple analytical model based on the latest LISA science requirements (SciRD) was used for the noise PSD. The functions $S_{\alpha}(f)$ are the sum of the analytic approximations to the instrumental noise curve in Ref.~\cite{2021_Babak_PSD} and galactic binary confusion noise in Eq.~4 of Ref.~\cite{Babak_EMRI_2017} scaled to a mission duration of 4 years. 
These are used to construct PSDs for each of the noise-orthogonal TDI channels $A,E$ and $T$. 

Within the search region set by the prior ranges, we use an iterative search strategy. 
Initially, the sampler is tasked with exploring the $\hat{L}_{N=N_{\rm max}}$ likelihood surface; this is expected to exhibit the broadest features which makes finding the peak possible. 
Once the optimizer/sampler has converged, the number of segments in the likelihood is reduced to $N=N_{\rm max}/4$; we find that reducing the number of segments by a factor of four at each stage is reasonably efficient for our fiducial source. 
The prior ranges on the phasing parameters are also reduced, with the new bound on each parameter calculated as the $98\%$ confidence interval of the 1D marginal posteriors from the previous stage. This simple approach shrinks the prior using progressively smaller hyper-rectangles.
The sampler is now tasked with exploring the new $\hat{L}_{N}$ likelihood surface with smaller $N$. 
This process is repeated, reducing $N$ and shrinking the prior ranges, until the sampler has explored the fully coherent $\hat{L}_{N=1}$ likelihood surface.
Sampling was performed using the \texttt{CPNest} nested sampling package \cite{Walter_2022_cpnest}.

The sequence of posteriors for a selection of the phasing parameters are shown in Fig.~\ref{fig:Inference_hierarchichal}. Constraints on these parameters improve throughout the iterative process. Fig.~\ref{fig:Inference_posterior_widths} shows the width of the 1D marginal posterior distributions for all parameters as a function of the segment number $N$, alongside the prior volume at different stages of the search.

For the BNS GW170817 observed in LIGO/Virgo, the sky position and time-of-merger parameters are not strongly impacted by the semi-coherent analysis (see Fig.~\ref{fig:BNS_posteriors}). 
These are generally referred to as extrinsic parameters and they don't impact the phasing of the GW signal. 
However, for SmBBHs in LISA this is not the case.
The time-to-merger parameter controls the frequency of the source at the start of observations and the sky position also affects the observed frequency via a periodic Doppler shifting caused by the detector motion. Therefore, posteriors on these parameters narrow during the search process (see Fig.~\ref{fig:Inference_hierarchichal}).

\section{Particle Swarm Optimization}\label{PySO}

\begin{figure*}[t!]
    \centering
    \subfloat{\includegraphics[width=0.9\textwidth]{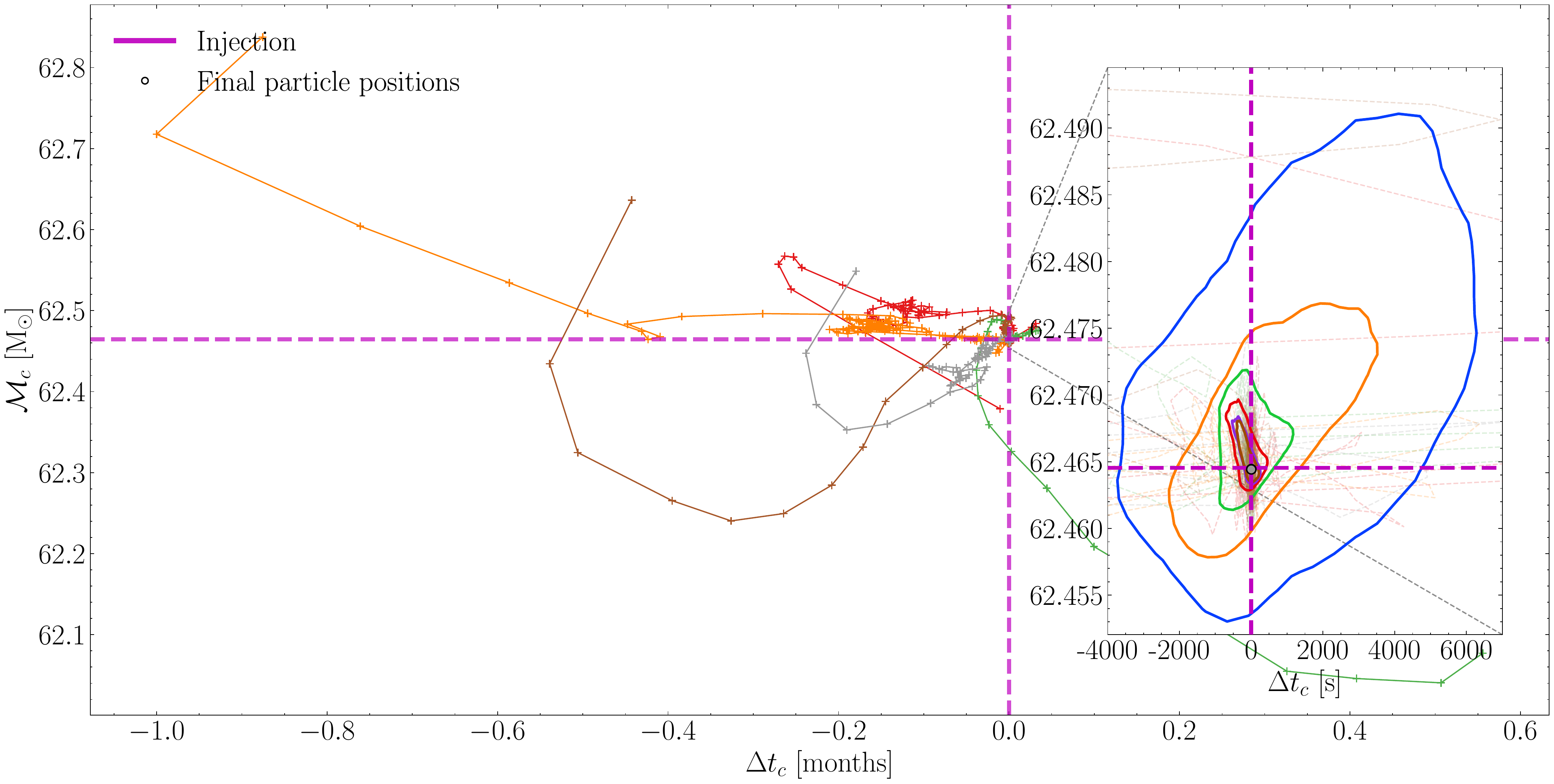}}
    
    \subfloat{\includegraphics[width=0.9\textwidth]{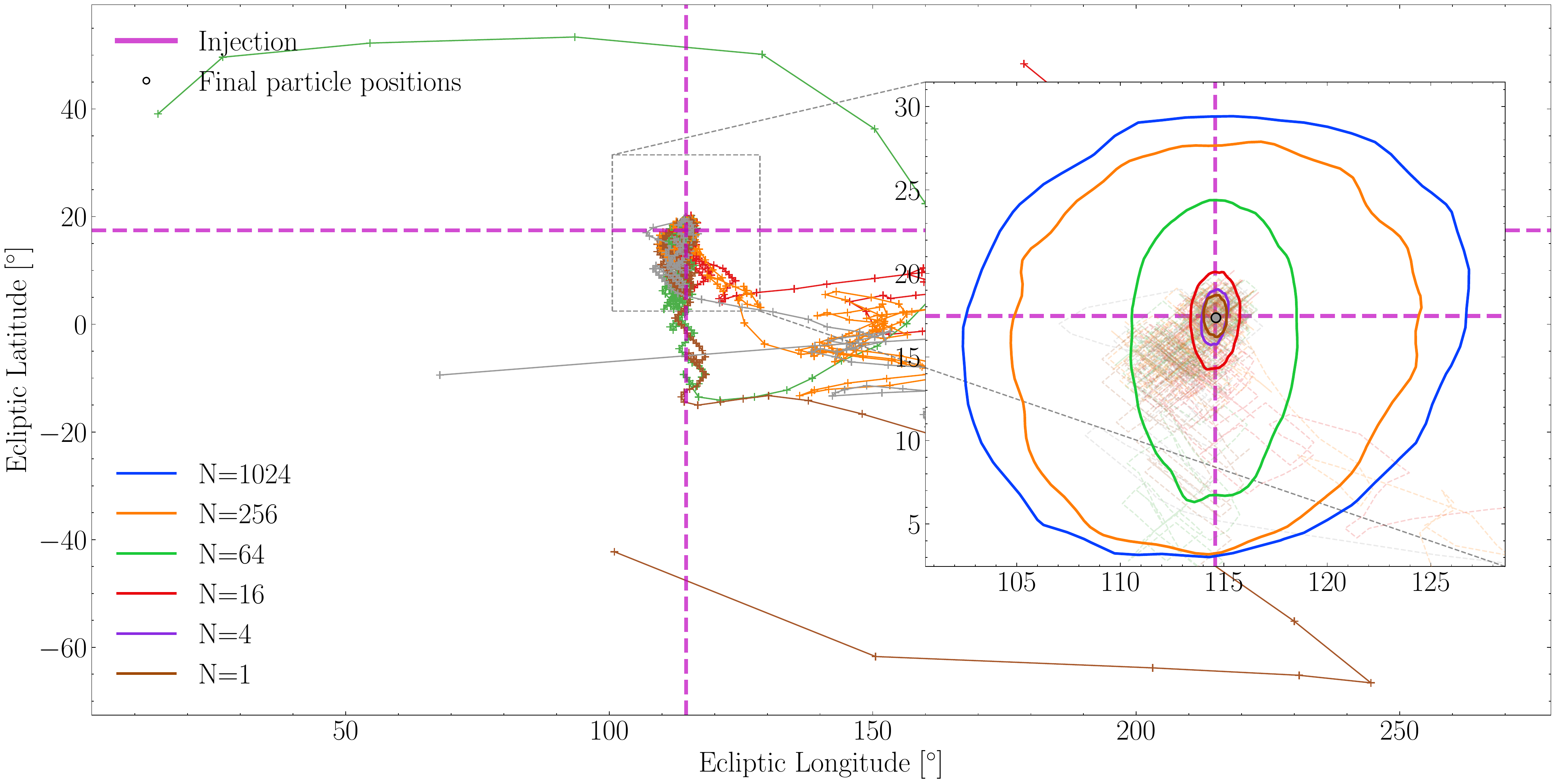}}
    \caption{Tracks in parameter space for 5 randomly selected particles, showing evolution throughout the PSO optimization on the (\emph{Top}) chirp mass, time to merger plane and (\emph{Bottom}) ecliptic longitude and latitude plane. Inset plots zoom in around the injection parameters, overlaying $90\%$ confidence intervals for posterior distributions obtained from analysis in Sec.~\ref{SmBBH_test_bench}.}
    \label{fig:PSO_pairplots}
\end{figure*}

\begin{figure*}[t!]
    \centering
    \includegraphics[width=\textwidth]{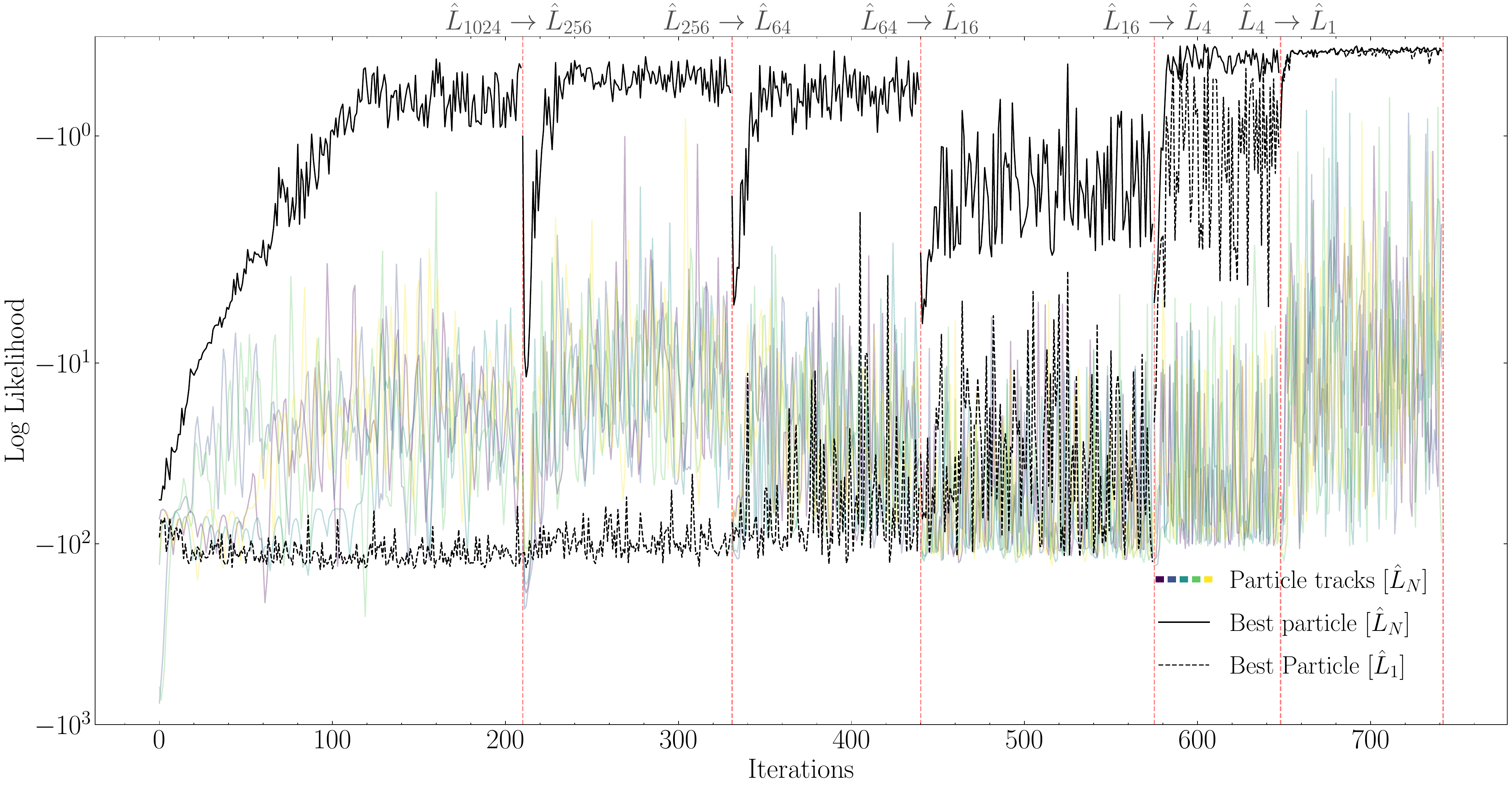}
    \caption{Log-likelihood evolution for the subset of particles plotted in Fig.~\ref{fig:PSO_pairplots} in solid lines. Dotted vertical lines indicate the iteration at which the optimizer switched to the next level in the hierarchical search. The colored lines show the track of the semi-coherent log-likelihood at the current number of segments, $\hat{L}_N$, for the 5 randomly selected particles (same particles as those shown in Fig.~\ref{fig:PSO_pairplots}), throughout the entire evolution of the swarm. 
    The black solid line shows the maximum semi-coherent log-likelihood of any particle in the swarm. (Note that the identity of the best particle changes repeatedly during the evolution, so the black line is not a particle track.)
    For comparison, the dotted black line shows the value of the coherent phase maximized log-likelihood $\hat{L}_{N=1}$ evaluated at the same parameter values as the solid black line.
    At early times the log-likelihood evaluated on the $N=1024$ segment for the best particle increases quickly, however the coherent likelihood stays relatively flat; this shows the impact of using the semi-coherent likelihood with peaks that are much broader and easier to find in parameter space.
    }
    \label{fig:PSO_LogL}
\end{figure*}

In the previous section we tested the iterative semi-coherent search by sampling the likelihoods; this is unnecessarily inefficient for a search. In this section a stochastic optimization algorithm is used to locate and track the peak of the semi-coherent likelihood with varying $N$ without wasting time exploring the tails of the likelihood distribution. Here, a Particle Swarm Optimizer (PSO) is used to do this, although the semi-coherent likelihood is general and can be used with any optimization algorithm.

PSO \citep{Kennedy_1995_PSO,1998_Shi_PSO} is a stochastic optimization algorithm which uses a swarm of a large number $N_p$ of particles to optimize an objective function over a high-dimensional parameter space. In this study we use it to optimize the semi-coherent log-likelihood, $\hat{L}_N$. 
Each of the $N_p$ particles in the PSO swarm has a position vector in parameter space that is updated at each iteration, $\theta^\mu_{p,i}$: the index $\mu\in\{0,1,\ldots,\mathrm{dim}(\theta)-1\}$ labels the components of the source parameter vector; the index $p\in\{0,1,\ldots, N_p-1\}$ labels the particles in the swarm; and the index $i\in\{0,1,\ldots\}$ labels the iteration of the algorithm. 
The rule for updating the positions at each iteration is
\begin{equation}
    \theta^{\mu}_{p,i+1} = \theta^{\mu}_{p,i} + v_{p,i}^{\mu},
\end{equation}
where $v^\mu_{p,i}$ is called the velocity and is calculated for each particle from the current state and past history of the whole swarm according to
\begin{equation}\label{eq:PSO}
    v_{p,i+1}^{\mu} = \mathrm{max}(\epsilon^{\mu}, |u^{\mu}_{p,i}|) \frac{u^{\mu}_{p,i}}{|u^{\mu}_{p,i}|} ,
\end{equation}
where
\begin{align}\label{eq:PSO2}
    u^{\mu}_{p,i} = & \Omega v_{p,i}^{\mu} +  \\ & \Phi_{P}(r_P)^{\mu}_{p,i} (\Psi_{p,i}^\mu-\theta_{p,i}^{\mu}) +\Phi_{G}(r_G)^{\mu}_{p,i}(\Xi_i^\mu-\theta_{p,i}^{\mu})\nonumber
\end{align}
and where $\Psi_{p,i}$ is the best (i.e.\ highest-likelihood) point visited by the $p^{\rm th}$ particle so far in the evolution of the swarm,
\begin{align}
    \Psi_{p,i}^\mu = \theta^\mu_{p,I} \;, \quad\mathrm{where} \quad I = \displaystyle\operatorname*{argmax}_{i'<i}\hat{L}_N(d|\theta_{p,i'}),
\end{align}
and where $\Xi_i$ is the best point visited by any particle
\begin{align}
    \Xi_i^\mu = \Psi^\mu_{P,i} \;, \quad\mathrm{where} \quad P = \displaystyle\operatorname*{argmax}_{p}\hat{L}_N(d|\Psi_{p,i}).
\end{align}
Eq.~\ref{eq:PSO2} is called the PSO velocity rule and has been widely used in the literature. 
It includes 3 terms: the $\Omega$ term  is called the \emph{inertia}; 
the $\Phi_P$ term is called the \emph{cognitive} term and acts to attract a particle back to the highest likelihood location that it has visited so far; 
and the $\Phi_G$ terms is called the \emph{social} term and acts to attract the particle towards to the highest likelihood location that any particle in the swarm has visited so far.
The quantities $(r_P)^\mu_{p,i}$ and $(r_G)^\mu_{p,i}$ are random numbers from $U[0,1]$ drawn independently for each particle, component and iteration.
 
Additional control over the behavior of the swarm can be achieved by varying the $\epsilon^{\mu}$ parameters in Eq.~\ref{eq:PSO}.
These control the minimum velocity any particle can have along a particular dimension of parameter space. 
Imposing a minimum velocity mitigates against premature convergence to local optima, which is a common problem for PSO methods \cite{2016_Larsen_PSO_premature_convergence} when optimizing multi-modal objective functions. 
Strictly speaking, imposing a minimum velocity in this way also prevents the swarm from ever converging, because the particles can never stop moving.
For this reason the $\epsilon^\mu$ parameters must be decreased near the end of the search. 
Similar \textit{velocity clamping} methods exist within the PSO literature, but these are usually aimed at restricting the maximum velocity of particles to prevent excessive exploration \cite{2000_Eberhart_Vmax,2015_Alhussein_Vmax}. 

Collectively, $\Omega$, $\Phi_P$, $\Phi_G$, and $\epsilon^\mu$ are referred to as the swarm hyperparameters.
The behavior of the swarm can be altered by varying the hyperparameters.
This allows us fine control over the rate of convergence and degree of exploration of the optimization algorithm, ideal for a semi-coherent search. 
Early in the search phase (high $N$) we want the swarm to focus on exploring the parameter space to ensure the peak is found. We want to avoid at all costs the swarm getting stuck and wasting time optimizing secondary peaks.
To achieve this the inertia, $\Omega$, is set high and the social weight, $\Phi_G$, is set low. Additionally, the minimum velocities, especially those in the most important phasing parameters are set to high values.
Late in the search (low $N$), when a broad likelihood feature has already been identified, the aim is to refine the parameter values by optimizing and converging on the peak.
To achieve this the inertia is decreased, and the social weight increased, and the minimum velocities are decreased.
These different behaviors are often referred to as \emph{exploration} and \emph{exploitation} in the PSO literature. 
Tab.~\ref{tab:PSO_params} in  appendix \ref{app:PySO} shows how all the PSO parameters vary throughout the search.

It is illustrative to compare PSO to a more well-known algorithm within the GW community, MCMC. PSO is similar in the sense that it is a stochastic algorithm where walkers move in a guided random way around parameter space. However, it differs in that it is an optimization, as opposed to a sampling algorithm and therefore tends to climb the likelihood surface without exploring the low-probability tails. PSO is also \emph{not} a Markov-Chain because the velocity rule incorporates `memory' of past positions through the $\Phi_P$ and $\Phi_G$ terms.

We use a swarm with $N_p=15000$ particles. The initial particle positions are drawn from the prior, with velocities in parameter $\theta^\mu$ drawn from the uniform distribution $U[-\Delta \theta/5,\Delta \theta/5]$ where $\Delta\theta$ is the prior range.
As in Sec.~\ref{SmBBH_test_bench}, we use $N_{\rm max}=1024$, and this likelihood surface is optimized over until the swarm is considered converged (see appendix \ref{app:PySO}). 
The optimizer is then configured for the next likelihood segment; dropping from $N_{\rm max}=1024$ to $N=256$. The particle positions are carried over from the final positions of the evolution at the previous level, mirroring the shrinking priors in the sampling. When dropping to a smaller number of segments the particle velocities are re-initialized by drawing from a zero-mean Gaussian distribution with a co-variance calculated from the final positions of all the particles in the previous segment.
The swarm hyper-parameters are also changed to gradually transition the swarm from exploratory to convergent behavior. 
This iterative process repeats until after $i_{\rm max}$ iterations the swarm has converged on the $N=1$ phase-maximized coherent likelihood. 
The final value of $\theta_{\rm best}\equiv \Xi_{i_{\rm max}}^\mu$ is our estimate of the best-fitting parameters and constitutes the main result of the search.

The colored lines in Fig.~\ref{fig:PSO_pairplots} shows particle tracks in selected parameters for $5$ randomly chosen particles from the swarm optimization throughout the evolution. 
Fig.~\ref{fig:PSO_LogL} shows the corresponding log-likelihood evolution tracks for these $5$ particles. 
Notice in Fig.~\ref{fig:PSO_LogL}, immediately after a step in the segment level, the log-likelihood curve (black solid line) drops significantly. Since the particle positions don't change as we move between segment levels, this drop in log-likelihood is due to the increased sensitivity to waveform phase of the semi-coherent likelihood with a smaller $N$. 
At each new level we observe that the function values get slightly worse in the first few iterations, before starting to improve. We attribute this to the particles initially exhibiting exploratory behavior due to the re-initialized random velocities.
The simple PSO algorithm implemented in this study localizes the parameters of the injected signal with good accuracy, specially in the phasing parameters which are of interest in the context of establishing narrow priors around the posterior bulk for parameter inference. For the fiducial signal considered here, the search has an execution time $\sim 15$ hours, although we obtain good estimates for phasing parameters such as chirp mass, time to merger and sky position parameters from the state of the optimizer at the end of the $1024$ segment level, within $\sim$ hours. Parameter estimation on the vanilla likelihood has a tractable computational cost when paired with priors derived from this PSO search result, successfully sampling the vanilla posterior for the fiducial source in $\sim 20$ hours using \texttt{CPnest}. 

\section{Future work and extension to the EMRI search problem} \label{sec:discussion}

In this work we demonstrated how a search using a likelihood with a variable level of coherence in conjunction with a particle-swarm-based optimizer can be used to find a SmBBH signal in mock LISA data. This demonstration has been performed in idealized data (i.e zero-noise, no gaps or glitches; see the discussion in Sec.~\ref{sec:introduction}). Further studies with realistic noise are needed to establish the sensitivity of this search at a fixed false alarm rate.

In this study we have demonstrated the search for a fiducial signal, searching within one `tile' in parameter space. Scaling this up and tiling the whole parameter space of interest is left for future work. We have also searched for only one source, a more realistic search would aim to find multiple sources, although this is not expected to be a major problem as it is unlikely there will be more than one source per search `tile' (see Tab.~1 of Ref.~\cite{2016_Kyutoku_number_of_detectable_SmBBH}). Precise tuning of our method (i.e. selecting the decreasing sequence of segment numbers between $N=N_{\rm max}$ and $N=1$, and the PSO hyperparameters) across parameter space is also left as future work. Finally, this method has not been tested with more complete waveforms that incorporate additional physics eg. eccentricity, spin-orbit precession etc, however this is not expected to be a major obstacle.

There are a number of similarities between SmBBH and EMRI signals observed by LISA. Designing and implementing a successful SmBBH search is likely to be extremely good preparation for the EMRI search problem. However, these two sources are astrophysically distinct from one another; LISA observes the early inspiral of SmBBH systems, these systems are in the regime $v \ll c$, meanwhile the late stage inspiral of EMRI system places them in the $v \lesssim c$ regime. Due to this and the extreme asymmetry in the mass ratio of the EMRI system, EMRI signals are comprised of dozens of frequency modes (also known as harmonics or \emph{voices}), each contributing a non-negligible fraction of the signal SNR. This is in stark contrast to SmBBH signals which are well described by just the leading order quadrupole frequency mode. The frequency evolution of each EMRI mode individually looks somewhat similar to a SmBBH inspiral. In addition to the many modes, EMRIs are generically expected to occur on eccentric and inclined orbits further complicating waveform generation \cite{Babak_EMRI_2017}; consequently EMRI waveforms are expensive to evaluate. Increasingly accurate and efficient time-domain EMRI waveforms are starting to become available \cite{2021_Katz_FastEMRIwaveforms} and these have started to be used for frequency-domain data analysis using a stationary phase approximation \cite{2023_Speri_FD_EMRI} for each frequency mode.
However, it is still not clear which EMRI waveforms will be available (and at what computational cost) for use in searches during the LISA mission.

The semi-coherent method developed here would need to be augmented to deal with the numerous modes in EMRI signals. The segmentation of the signal done here in the frequency domain applied to an EMRI would analyze different modes at different times.
Therefore, it might be necessary to analyze each mode individually with this type of frequency-domain semi-coherent method. An investigation of this multi-mode, semi-coherent analysis is left to future work.

Unlike the SmBBH likelihood, the EMRI likelihood surface exhibits an extreme degree of multi-modality, this was recently studied in detail within Ref.~\cite{Chua_2021}. This likelihood has many spurious secondary peaks of comparable height to the primary likelihood peak around the true source parameters. It is worth emphasizing these peaks \emph{do not} originate from noise artifacts within the data but rather due to alignment of frequencies (and frequency derivatives) between waveforms evaluated at different points in parameter space. Ref.~\cite{Chua_2021} concluded there is not a strong relationship between the height of the secondary likelihood peaks with the euclidean distance between secondary peak parameters and the injection, hundreds of secondary peaks were found in the likelihood surface relatively close to the injection parameters. This makes the EMRI search even more challenging than the SmBBH case.

PSO is a highly flexible algorithm which can be tuned for this extremely degenerate likelihood surface. 
The velocity rule can be easily adapted to split a swarm into multiple sub-swarms. We propose the use of this multi-population particle swarm optimizer to explore this degenerate likelihood surface. Similar variants of particle swarm optimization for multi-modal objective functions have been studied in Refs.~\citep{Parrot_2006_SPSO,Kennedy_2000_SPSO}. Each swarm will be assigned to a peak and optimize across them in parallel, prioritizing the `best' peaks. Similar methods are used in Ref.~\cite{Petiteau_2010_GA} to sort through multiple optima, using genetic algorithms to search for massive binary BH mergers in mock LISA data.

The initial configuration of such a multi-population swarm would be similar to that of the vanilla version presented in this study, comprised of one exploratory swarm with a greater weight in $\Phi_P$, causing the particles to cluster around the large number of optima. At the end of the first segment level, a clustering algorithm such as that suggested in Ref.~\cite{Chua_2021} will be applied, clustering the single population swarm into multiple populations, each exploring one optima. Over the course of the next segment level, each population will be treated as an individual swarm, optimizing over its own maxima. Clustering will be conducted at the end of each segment level, terminating swarms that are exploring sub-dominant peaks according to some `veto' criteria, such as that suggested in Ref.~\cite{Chua_2021} and redistributing the particles to the other swarms. One such `veto' function, is proposed in Ref.~\cite{Chua_2022_veto},
calibrated to suppress secondary maxima that arise due to phase matching of the dominant frequency mode. 

In conclusion the semi-coherent search is a promising avenue for broadband, chirping sources which undergo many orbits in the LISA frequency band and warrants further investigation into its efficacy as a search pipeline for SmBBH and EMRI sources.

\vspace{0.4cm}
\begin{acknowledgments}
    We would like to thank all the other developers of the \texttt{Balrog} software package. 
    We also thank Graham Woan for helpful discussions on semi-coherent methods and Riccardo Buscicchio and Eliot Finch for helpful comments on the manuscript. The computations described in this paper were performed using the University of Birmingham's BlueBEAR HPC service. D. B. is supported by the STFC and the School of Physics and Astronomy at the University of Birmingham. D. B. also acknowledges the support of the Google Cloud Research Credits program, Grant No. 289387648. C. J. M. acknowledges the support of the UK Space Agency, Grant No. ST/V002813/1.
\end{acknowledgments}


\bibliographystyle{apsrev4-1}
\bibliography{bibliography}


\appendix

\section{Tempering vs semi-coherent}\label{tempering_comparison}

A tempered version of a probability distribution $p$ with temperature $T$ is the new probability distribution  
\begin{align}\label{eq:Tempering}
    \Tilde{p} = p^{1/T}.
\end{align}
In its application to GW data analysis, a tempered posterior distribution aids the exploration of stochastic sampling algorithms and mitigates against sampling chains getting stuck in secondary maxima for multi-modal likelihood surfaces. Considering the simplest sampling algorithm, the Metropolis Hastings MCMC sampler, a proposal for a walker at position $\theta_i$ to move to $\theta_j$ is accepted with probability $\alpha = p(\theta_j)/p(\theta_i)$. In the situation where the walker is initially in a secondary maxima, it is unlikely to step out and into another well-separated global maxima. Parallel tempering broadens peaks by raising the log-likelihood floor, increasing the probability the walker will be able to jump between maxima and therefore explore the multi modalities in the surface. From Eq.~\ref{eq:Tempering} it can be seen that tempering is a re-scaling of the log-likelihood.

We contrast this with the semi-coherent method which broadens log-likelihood around the injection and combines secondary peaks smoothly, \emph{removing} variations in the log-likelihood surface. Whereas the tempering approach \emph{suppresses} variations in the log-likelihood by raising the troughs, however it does not remove the secondary peaks, just suppresses differences between the peaks and troughs. These suppressed variations result in a surface that is easier for a probabilistic sampling algorithms to explore, however the secondary peaks still exist in the surface, making it difficult for optimization algorithms (such as those conventionally used in the search/prior localization phase) to explore this surface. Meanwhile the semi-coherent approach results in a smooth surface around the injection, \emph{removing} variations in the log-likelihood, this approach is much better suited for exploration by an optimizer. 

\begin{figure*}[h]
    \centering
    \includegraphics[height=0.85\textheight]{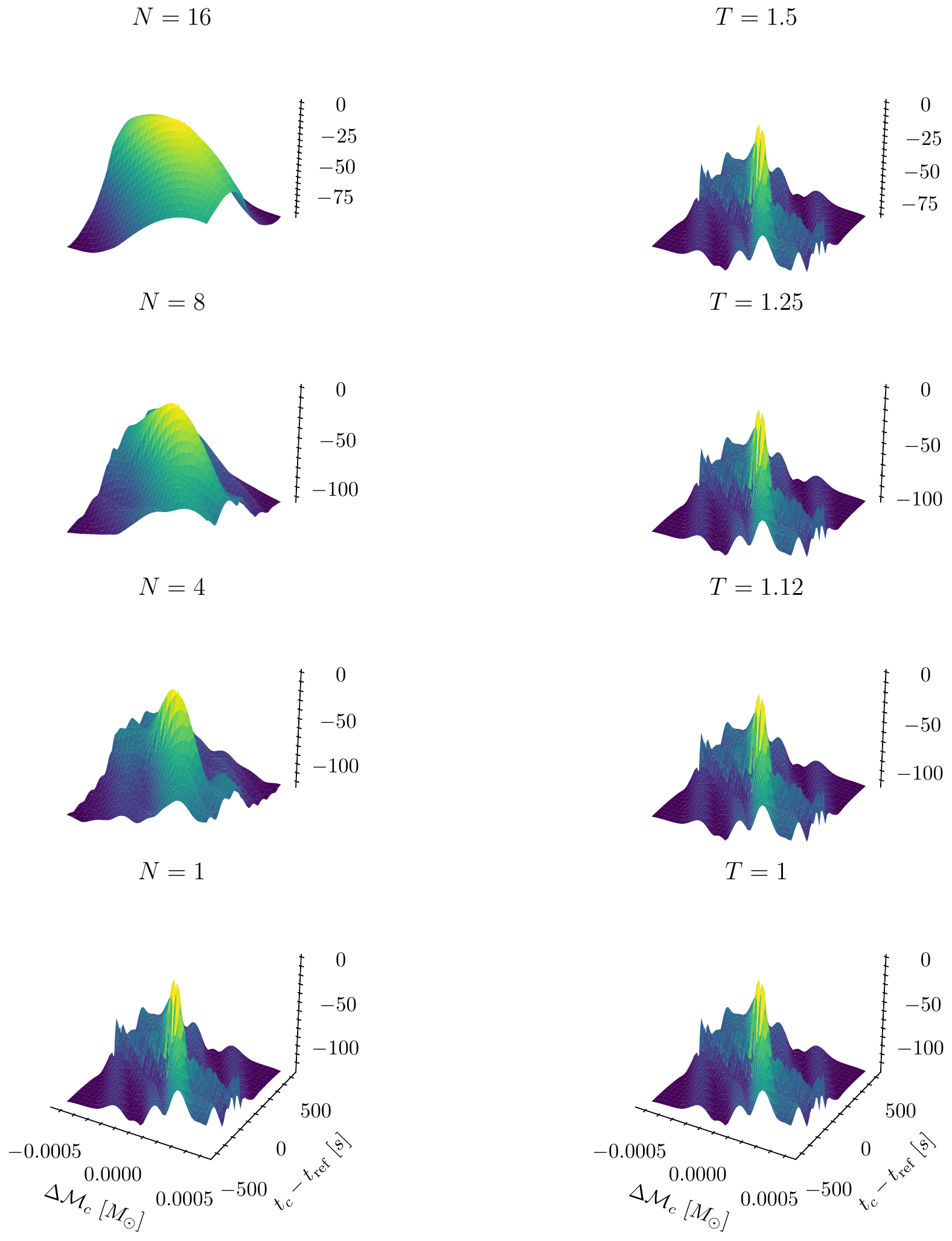}
    \caption{Figure illustrating the difference between a tempered log-likelihood surface, $T^{-1}\mathrm{log}L(d|\theta)$, (\emph{right column}) with variable temperature $T$ and a semi-coherent log-likelihood, $\mathrm{log}\hat{L}_{N}$ (\emph{left column}).
    The likelihood surfaces plotted here are for the SmBBH LISA signal described in Sec.~\ref{SmBBH_test_bench} and Tab.~\ref{tab:Injection} (plotted as a function of the chirp mass and time to merger parameters with all other parameter fixed to their true values) although the trends shown are generic.
    Increasing $N$ in the semi-coherent likelihood raises the floor of the likelihood surface, decreasing the peak-to-trough range of log-likelihood values. It also has the effect of congealing secondary-maxima together; the complicated structure of peaks troughs and ridges seen in the coherent ($N=1$) are completely absent in the top $N=16$ plot. 
    In contrast, the tempered log-likelihood surface is simply a re-scaled version of the vanilla log-likelihood, this has the effect of only raising the log-likelihood floor. 
    The values of $T$ plotted here were chosen so that the range of log-likelihood values in the region plotted is similar between the left and right columns of plots. 
    }
    \label{fig:Semi_coherent_vs_tempering}
\end{figure*}

A comparison between the tempered and semi-coherent likelihoods is shown in Fig.~\ref{fig:Semi_coherent_vs_tempering}. 

\section{PSO configuration} \label{app:PySO}
The PSO search used in this study used $N_p=15000$ particles, with log-likelihood evaluations parallelized across 20 CPU cores. The optimization of each hierarchical segment (i.e.\ value of $N$) was allowed a maximum of $250$ iterations, however the PSO swarm could stop and move to the next log-likelihood (i.e.\ $N \rightarrow N/4$) at iterations below this if they met the convergence criteria. In this study we set the convergence criteria as the absence of any significant improvement (with a tolerance of $0.01$) in the best swarm function value in the last $50$ iterations. 
Tab.~\ref{tab:PSO_params} shows the hyperparameters used for each stage of the hierarchical search. 

\begin{table*}
\caption{\label{tab:PSO_params}
    The PSO hyperparameters used throughout the analysis.
    In the first stage the PSO optimizes the semi-coherent likelihood with $N_{\rm max}=1024$ segments using the hyperparameter settings in the first row of the table.
    In subsequent stages the number of segments is progressively reduced, e.g.\ to $N=256$ in the second row, and the the hyperparameters changed accordingly.
    The inertial weight $\Omega$ is reduced throughout the analysis and the social weight $\Phi_G$ is increased to gradually transition the behavior of the swarm from exploration to exploitation. 
    The minimum velocities for each parameter $\epsilon^\mu$ are also reduced during the run which also helps with the transition from exploration to exploitation. (The $\epsilon^\mu$ parameters are dimensionfull and have the same units as the corresponding parameters in Tab.~\ref{tab:Injection}.)}
\begin{ruledtabular}
\begin{tabular}{c|ccc|ccccccccccc}
    \textrm{Segment} &
    \textrm{$\Omega$} & 
    \textrm{$\Phi_P$} & 
    \textrm{$\Phi_G$} &
    \textrm{$\epsilon^{\lambda}$} & 
    \textrm{$\epsilon^{\sin \beta}$} & 
    \textrm{$\epsilon^{t_c}$} & 
    \textrm{$\epsilon^{\mathcal{M}_c}$} & 
    \textrm{$\epsilon^{\delta \mu}$} & 
    \textrm{$\epsilon^{\chi_1}$} & 
    \textrm{$\epsilon^{\chi_2}$} & 
    \textrm{$\epsilon^{\phi_{\rm left}}$} & 
    \textrm{$\epsilon^{\phi_{\rm right}}$} &
    \textrm{$\epsilon^{\sqrt{A_{\rm left}}}$} & 
    \textrm{$\epsilon^{\sqrt{A_{\rm right}}}$} \\
    \colrule
    $1024$ & $0.6$ & $0.2$ & $0.2$ & $10^{-2}$ & $10^{-2}$ & $50$ & $10^{-2}$ & $10^{-1}$ & $10^{-1}$ & $10^{-1}$ & $10^{-1}$  & $10^{-1}$ & $5 \times 10^{-6}$ & $5 \times 10^{-6}$ \\
    $256$  & $0.5$ & $0.2$ & $0.3$ & $10^{-2}$ & $10^{-2}$ & $50$ & $10^{-2}$ & $10^{-1}$ & $10^{-1}$ & $10^{-1}$ & $10^{-1}$  & $10^{-1}$ &  $10^{-6}$ &  $10^{-6}$ \\
    $64$   & $0.4$ & $0.2$ & $0.4$ & $10^{-2}$ & $10^{-2}$ & $50$ & $10^{-2}$ & $5 \times 10^{-2}$ & $5 \times 10^{-2}$ & $5 \times 10^{-2}$  & $10^{-1}$  & $10^{-1}$ & $10^{-6}$ & $10^{-6}$ \\
    $16$   & $0.3$ & $0.2$ & $0.5$ & $10^{-2}$ & $10^{-2}$ & $50$ & $10^{-2}$ & $10^{-2}$ & $5 \times 10^{-2}$ & $5 \times 10^{-2}$        & $10^{-1}$  & $10^{-1}$ & $10^{-6}$ & $10^{-6}$ \\
    $4$    & $0.2$ & $0.2$ & $0.6$ & $10^{-3}$ & $10^{-3}$ & $25$ & $10^{-4}$ & $10^{-2}$ & $10^{-2}$ & $10^{-2}$ & $10^{-1}$  & $10^{-1}$ & $10^{-6}$ & $10^{-6}$ \\
    $1$    & $0.2$ & $0.2$ & $0.6$ & $10^{-4}$ & $10^{-4}$ & $5$  & $10^{-5}$ & $10^{-3}$ & $10^{-3}$ & $10^{-3}$ & $10^{-1}$  & $10^{-1}$ & $10^{-6}$ & $10^{-6}$ \\
\end{tabular}
\end{ruledtabular}
\end{table*}

We stress the PSO configuration and ladder of segments used in this study are tuned empirically and while it is sufficient at the level of a proof-of-concept study, further work is needed to provide concrete suggestions. 
It is likely such configurations will be source dependent, e.g.\ SmBBH systems will possibly have a very different optimal PSO configuration and segment ladder to EMRIs.

\section{Parameter transforms}\label{app:Param_transforms}
The following are the definitions of sampling parameters used in sections \ref{SmBBH_test_bench} and \ref{PySO} for the SmBBH LISA analysis.

\begin{equation}\label{eqn:chirp_mass}
    \mathcal{M}_c = \frac{(m_1 m_2)^{\frac{3}{5}}}{(m_1+m_2)^{\frac{1}{5}}}
\end{equation}

\begin{equation}\label{eqn:delta_mu}
    \delta \mu= \frac{m_1-m_2}{m_1+m_2}
\end{equation}

\begin{equation}\label{eqn:A_LR}
    \sqrt{A_{\mathrm{left/right}}}  = \sqrt{\frac{1}{2 d_L}}\bigg(1\pm\cos(\iota)\bigg)
\end{equation}

\begin{equation}\label{eqn:phi_LR}
    \phi_{\mathrm{left/right}}  = \phi \mp 2\psi
\end{equation}

\section{Quadrature methods for large prior and noisy analyses}\label{Clensahw_curtis_comments}

The integral of a function $f(x)$ in the range $[x_{\rm min},x_{\rm max}]$ can be approximated using the Riemann sum
\begin{align}
    \int_{x_{\rm min}}^{x_{\rm max}} f(x) \mathrm{d}x \approx \sum_{i=0}^{N_{R}-1} f(x_i) \Delta x
\end{align}
where $\Delta x=(x_{\rm max}-x_{\rm min})/N_{R}$ and $x_i = x_{\rm min} + i \Delta x$. The set of points $\{x_i,i=0,1,\ldots,N_{R}\}$ constitute a discrete, uniform grid over which the integral is computed. 
Alternatively, this integral can be approximated using a quadrature rule;
\begin{align}
    \int_{x_{\rm min}}^{x_{\rm max}} f(x) \mathrm{d}x \approx \sum_{i=0}^{N_{Q}-1} w(x_i) f(x_i)
\end{align}
where the irregularly spaced $x_i$ nodes are located at the roots of (suitably rescaled) Chebyshev polynomials and where $i$ indexes the $N_{\rm Q}$ quadrature nodes. Quadrature numerical integration methods can achieve can typically achieve a given accuracy of approximation using a far smaller number of nodes ($N \ll N_{\rm Riemann}$) for smooth integrand functions $f$. The weights $w(x_i)$ only depend on the limits and can be pre-computed. These methods are commonly based on interpolating $f(x)$ over the domain using interpolation functions (Chebyshev polynomials in this case) which have known analytic integrals used to generate $w(x_i)$.

In this study, Clenshaw-Curtis quadrature is used to evaluate the semi-coherent likelihood in Eq.~\ref{eq:def_semico_L_2}. However, such methods are limited to integrands which are smooth over the integration domain due to the interpolation function usually being smooth. An example of smooth function where quadrature integration performs well is the $\left<h|h\right>$ term in the log-likelihood (this is also equal to the squared SNR). Consider a signal $h(\theta) = A(f)e^{i\phi(f)}$, the squared SNR is given by
\begin{align} \label{eq:app_hh}
    \bra{A(f)e^{i\phi(f)}}\ket{A(f)e^{i\phi(f)}}\sim \int |A(f)|^2 \mathrm{d}f
\end{align}
where $A(f)$ is a smooth, slowly varying function over frequency (for clarity, we have omitted the factor of of $S(f)$ and other terms that do not effect the argument here from Eq.~\ref{eq:app_hh}). Thus this integral is well approximated by quadrature rules. 
Instead consider the case we have some data $d=h(\theta_*)=\Tilde{A}(f)e^{i\Tilde{\phi}(f)}$, where the parameters $\theta_*$ are the injected parameters of the source and we are performing a zero-noise injection. Consider the $\left<h|d\right>$ term in the log-likelihood when $\theta \neq \theta_*$. The inner product is now
\begin{align}
    \bra{h}\ket{d}\sim \int A(f)\Tilde{A}^{\dagger}(f)e^{-i(\phi(f)-\Tilde{\phi}(f))} \mathrm{d}f .
\end{align}
The term $e^{-i(\phi(f)-\Tilde{\phi}(f))}$ introduces oscillations into the integrand across the frequency domain. Assuming $\theta - \theta_*$ is small, $\phi(f)-\Tilde{\phi}(f)$ will likely be small, in this regime quadrature rules are still valid. This is the case for parameter estimation of broadband signals shown in Refs.~\citep{2022_Klein_Multiband,Buscicchio_2021,Pratten_2022} where narrow priors are used. This method of evaluating the likelihood is only valid when $\theta-\theta_*$ is small. This sets a maximum size on each search tile. If the waveform model is evaluated at a location far from the injection in parameter space, $\phi(f)$ and $\Tilde{\phi}(f)$ can be very different and the integrand oscillates rapidly over the frequency domain, quadrature rules are no longer valid. Note however that oscillatory integrands usually cancel to give small integrals. Introducing any source of rapid oscillations into the integrand will result in the quadrature approach failing. For example if the data $d$ contains noise (which is discontinuous between frequency bins), the integrand for $\bra{h}\ket{d}$ is oscillatory and thus cannot be evaluated using quadrature integration. This issue with the likelihood computation was highlighted in Ref.~\cite{2021_Marsat_integral_problems}. We have verified the quadrature grid used in this study produces likelihoods that are sufficiently faithful to those evaluated on the FFT grid, for parameters within our prior bounds, in the zero-noise scenario. 

Such problems with highly oscillatory integrals are not unique to quadrature rules, the prevalent alternative to quadrature rules used in frequency domain analyses for mock LISA data is heterodyning/relative binning of the likelihood. This uses a template waveform (on the FFT frequency grid), and expresses a waveform model evaluated at another location in parameter space as a slowly varying difference between the model and the template waveform. This method is also limited in the distance one can travel in parameter space from the template before the model waveform becomes inaccurate \cite{Cornish_2021}. Other methods evaluating this sort of oscillatory integral have been proposed, see Refs.~\cite{Chua_2021,2021_Marsat_integral_problems}. While it is not yet clear which method will be used for real, noisy LISA data, GPU hardware accelerated likelihoods are a promising avenue \cite{2020_Katz_BBHX,2023_Speri_FD_EMRI} which circumvents the previously discussed problems by generating waveforms directly on the FFT frequency grid.

\end{document}